\documentclass{aa}

\usepackage{graphicx}

\begin{document}


\title{Neutron star mergers versus core-collapse supernovae as dominant
       r-process sites in the early Galaxy}

\titlerunning{Neutron star mergers vs. SNe~II as dominant r-process sites}

\author{D.~Argast~\inst{1,2} \and M.~Samland~\inst{2} 
        \and F.-K.~Thielemann~\inst{1} \and Y.-Z.~Qian~\inst{3}}

\authorrunning{D.~Argast et al.}

\institute{Institut f\"ur Physik der Universit\"at Basel,
           Klingelbergstrasse 82, CH-4056 Basel, Switzerland
\and       Astronomisches Institut der Universit\"at Basel,
           Venusstrasse 7, CH-4102 Binningen, Switzerland
\and       School of Physics and Astronomy, University of Minnesota, 
           Minneapolis, Minnesota 55455, USA}

\offprints{D.~Argast, \\\email{argast@quasar.physik.unibas.ch}}

\date{Received ... / Accepted ...}


\abstract{The astrophysical nature of r-process sites is a long standing
mystery and many probable sources have been suggested in the past, among them
lower-mass core-collapse supernovae (in the range $8 - 10 \,
\mathrm{M}_{\sun}$), higher-mass core-collapse supernovae (with masses $\ge 20
\, \mathrm{M}_{\sun}$) and neutron star mergers.  In this work, we present a
detailed inhomogeneous chemical evolution study that considers for the first
time neutron star mergers as major r-process sources, and compare this
scenario to the ones in which core-collapse supernovae act as dominant
r-process sites. We conclude that, due to the lack of reliable iron and
r-process yields as function of progenitor mass, it is not possible at present
to distinguish between the lower-mass and higher-mass supernovae scenarios
within the framework of inhomogeneous chemical evolution. However,
neutron-star mergers seem to be ruled out as the \emph{dominant} r-process
source, since their low rates of occurrence would lead to r-process enrichment
that is not consistent with observations at very low metallicities.
Additionally, the considerable injection of r-process material by a single
neutron-star merger leads to a scatter in [r-process/Fe] ratios at later times
which is much too large compared to observations.
}

\maketitle


\section{Introduction}

The principal production mechanisms of heavy elements beyond the iron-peak are
known since the classical paper by Burbidge, Burbidge, Fowler \& Hoyle
(B$^2$FH, \cite{b2fh}): A major fraction of these elements is formed by slow
or rapid capture of neutrons on seed nuclei. These nucleosynthesis channels
are denoted as the s-process, if the capture of neutrons occurs on timescales
slower than the $\beta$-decay timescales of the newly formed nuclei, or as
the r-process, if neutron capture operates on timescales much shorter than the
relevant $\beta$-decay timescales. Yet, although the physical
requirements for the occurrence of r-process nucleosynthesis are well
understood (see e.g. Pfeiffer et al.~\cite{pf01}), the astrophysical nature of
the dominant r-process site is still unknown.

Ultra metal-poor halo stars were formed when the s-process had little time to
make any significant contribution to the interstellar medium (ISM). The
abundance pattern of neutron capture elements heavier than Ba ($Z \ge 56$) in
these stars matches the scaled solar system r-process abundances remarkably
well. This suggests that, contrary to s-process nucleosynthesis, the synthesis
of r-process elements started early in galactic evolution and that the
r-process is robust for elements heavier than Ba, i.e. originates from a
single astrophysical site or at least occurs under well defined physical
conditions (e.g. Sneden et al. \cite{sn00}; Westin et al.  \cite{we00}; Cowan
et al. \cite{co02}; but see also Hill et al. \cite{vh02}). On the other hand,
lighter neutron capture elements ($30 < Z < 56$) in ultra metal-poor halo
stars show significant deviations from the scaled solar system r-process
abundance curve, indicating the possible existence of a second r-process
source. The diversity of r-process sources was first proposed based on
meteoritic data, which seem to require two r-process sources operating on
different timescales, namely $\sim 10^7$ years for the production of $Z \ge
56$ nuclei and $\sim 10^8$ years for $Z < 56$ nuclei (Wasserburg et
al. \cite{wa96}). In principle, these two r-process sites could either be two
types (frequent and less common) of core-collapse supernova (SN~II) events
(Wasserburg \& Qian \cite{wa00}; Qian \cite{qi00}; \cite{qi01}), or a mixture
of SNe and neutron star mergers (Rosswog et al. \cite{ro99}; \cite{ro00}). On
the other hand, Cameron (\cite{cm01}) showed that it is possible to reproduce
the abundance distribution of both light and heavy neutron capture elements
in different regions of a single source, namely the accretion-disk and jet
forming near the proto-neutron star in SN~II events. However, the evidence
that the r-process beyond Ba (at least in the range $56 \le Z \le 72$) is
robust is very convincing and we will use this property in the following to
gain some insight into the enrichment of the Galaxy with r-process elements
beyond Ba.

A number of possible astrophysical sites responsible for the robust r-process
were put forth in the past, most of them linked to the violent death of
massive stars in supernova events. Wheeler et al. (\cite{wh98}) suggested
that, during the prompt explosion of a massive star in the range $8 - 10 \,
\mathrm{M}_{\sun}$, physical conditions in the innermost mass layers are
sufficiently extreme for the r-process to work and that the amount of
r-process matter ejected may be consistent with observed Galactic r-process
abundances. This is in agreement with detailed r-process calculations based on
an artificially induced prompt explosion of an $11 \, \mathrm{M}_{\sun}$
progenitor (Sumiyoshi et al. \cite{sm01}).  However, there are major
objections to the prompt explosion mechanism from detailed studies by Bruenn
(\cite{bru89a}, \cite{bru89b}).  Optimistically speaking, if \emph{prompt}
explosions of massive stars may occur in reality, lower-mass progenitors are
the only ones that show any hope to do so (e.g. Wheeler et al. \cite{wh98};
Sumiyoshi et al. \cite{sm01}; but see also Liebend\"orfer et al. \cite{li01}).

On the other hand, neutrino-driven winds from nascent neutron-stars were also
proposed as a promising site for r-process nucleosynthesis (e.g. Woosley \&
Hoffman \cite{wo92}; Takahashi et al. \cite{tk94}; Woosley et al. \cite{wo94};
Qian \& Woosley \cite{qi96}; Thompson et al. \cite{tp01}; Wanajo et
al. \cite{wn01}; Terasawa et al. \cite{te02}). Following the \emph{delayed}
explosion of a very massive star ($\ge 20 \, \mathrm{M}_{\sun}$), neutrinos
diffusing out of the contracting proto-neutron star heat and ablate material
from its surface and a neutrino-driven wind develops. Such winds may exhibit
sufficiently high entropies and/or sufficiently short dynamical timescales for
the r-process to occur.  However, r-process yields consistent with observed
r-process abundances in stars may be obtained in this scenario only for
extreme assumptions such as massive neutron stars of $2 \, \mathrm{M}_{\sun}$
or more, which makes this not a very likely scenario.  Furthermore, it seems
questionable whether the high entropies required to reproduce the solar system
r-process signature (e.g. Meyer \& Brown \cite{mb97}; Freiburghaus et
al. \cite{fr99a}) can be provided by Type~II supernovae (Witti et
al. \cite{wi94}; Qian \& Woosley \cite{qi96}; Thompson et al. \cite{tp01}).

It should be noted that the r-process scenarios associated with core-collapse
supernovae discussed above are affected by considerable theoretical
uncertainties since the physics of these events is not really understood
(e.g. Rampp \& Janka \cite{ra00}; Liebend\"orfer et al. \cite{li01}).
Recently, Freiburghaus et al. (\cite{fr99b}) presented for the first time
detailed r-process calculations for neutron star mergers (NSM). Coalescing
neutron stars potentially can provide in a natural way the large neutron
fluxes required for the build-up of heavy elements through rapid neutron
capture. This scenario was subsequently explored and refined by Rosswog et
al. (\cite{ro99}; \cite{ro00}). Their calculations showed that a few times
$10^{-3} - 10^{-2} \, \mathrm{M}_{\sun}$ of r-process matter might be ejected
in a merger event. This amount is significantly larger than the typical
$10^{-5}-10^{-6} \, \mathrm{M}_{\sun}$ of r-process material thought to be
ejected in each core-collapse SN event (Cowan et al. \cite{co91}; Woosley et
al. \cite{wo94}; Qian \cite{qi00}; Wanajo et al. \cite{wn01}). Since the rate
of NSM in the Galaxy is significantly lower than that of Type~II SNe
(e.g. Tamman et al. \cite{tm94}; Belczynski et al. \cite{bl02}), either of
these two sources may account for the total amount of r-process matter in the
Galaxy (Qian \& Woosley \cite{qi96}; Rosswog et al. \cite{ro99}; Rosswog \&
Davies \cite{ro02}; Thielemann et al. \cite{th02}).

However, Qian (\cite{qi00}) argued that if NSM were a major r-process source,
the low coalescence rate of binary compact objects would prevent any
correlation between abundances of r-process elements and iron, which is
clearly in disagreement with observed r-process abundances in stars more
metal-rich than [Fe/H] $\sim -2.5$.  In addition, the large amount of
r-process ejecta from NSM would lead to r-process abundances in metal-poor
halo stars that would be too large compared to observations.  These arguments
deserve closer examination, especially since no detailed chemical evolution
calculations exist to date that assume NSM to be the major r-process
source. Chemical evolution studies that treat SN with either lower-mass ($8 -
10 \, \mathrm{M}_{\sun}$) or higher-mass ($\ge 20 \, \mathrm{M}_{\sun}$)
progenitors as the major r-process source were carried out by Ishimaru \&
Wanajo (\cite{is99}), Travaglio et al. (\cite{tr99}) and Tsujimoto et
al. (\cite{ts00}). Unfortunately, each of these studies favoured different
r-process sites: lower-mass SNe (Travaglio et al. \cite{tr99}), higher-mass
SNe (Tsujimoto et al. \cite{ts00}), or both (Ishimaru \& Wanajo \cite{is99})
are acceptable from the viewpoint of chemical evolution.

The purpose of this work is twofold. First, neutron star mergers are included
as major r-process site in detailed chemical evolution calculations for the
first time. Second, we compare the r-process enrichment of the ISM under the
assumption that neutron star mergers, lower-mass SNe~II or higher-mass SNe~II
are the dominant r-process sites in the framework of inhomogeneous chemical
evolution. In Sect.~\ref{ice.model}, the inhomogeneous chemical evolution
model is presented. The impact of SNe~II and NSM as r-process sites on
Galactic chemical evolution is studied in Sect.~\ref{enrichment} and
conclusions are presented in Sect.~\ref{con}.


\section{The chemical evolution model}
\label{ice.model}

The large scatter seen in the abundances of neutron capture elements in
metal-poor halo stars is generally attributed to local chemical
inhomogeneities of the interstellar medium during the earliest stages of
Galaxy formation and enrichment (e.g. McWilliam et al. \cite{mw95a},
\cite{mw95b}; Ryan et al. \cite{ry96}; Burris et al. \cite{br00}; Mishenina
\& Kovtyukh \cite{mi01}). Several authors developed chemical evolution models
(Ishimaru \& Wanajo \cite{is99}; Raiteri et al. \cite{ri99}; Tsujimoto et
al. \cite{ts99}, \cite{ts00}; Argast et al. \cite{ar00}; Travaglio et
al. \cite{tr01}, Fields et al. \cite{fi02}) that investigate the impact of
local chemical inhomogeneities on the enrichment of the ISM and the scatter in
element abundances. Although the methods employed by these authors differ
significantly, the model results are in qualitative agreement. The notion of
inhomogeneous chemical evolution therefore seems to be well established.

In the present work we use the stochastic chemical evolution model presented
in Argast et al. (\cite{ar00}, hereafter Paper~I) to investigate the
enrichment of the ISM with r-process elements. In the following, we give a
short summary of the model and discuss some major changes (mass infall, star
formation rate, SN~Ia events and r-process sites) that were implemented for
this work.

\subsection{Basic ingredients}
\label{ice}

We model the chemical enrichment of the halo ISM within a volume of (2.0
kpc)$^3$, down to a resolution of (50 pc)$^3$. Primordial matter is assumed to
be falling into this volume, following an infall law of the form
\begin{equation}
\dot{M}\left( t \right) = a \cdot t^b \cdot \exp(-t/\tau).
\end{equation}

Instead of specifying parameters $a$ and $b$ directly, we use the fact that
with this description the time of maximal infall, $t_{max}$, is given by
\begin{equation}
t_{max} = b \cdot \tau,
\end{equation}
and that the total mass, $M_{tot}$, falling into the volume is
\begin{equation}
M_{tot} = \int_0^{t_{end}} a \cdot t^b \cdot \exp(-t/\tau) \, dt,
\end{equation}
where $t_{end}$ is the age of the system. The infall law is therefore fully
described by the parameter set $\left\{ M_{tot}, \tau, t_{max}, t_{end}
\right\}$.

A crucial ingredient of chemical evolution models is the treatment of star
formation (SF). Unfortunately, it is also one of the least understood. In this
work we adopt a simple SF description based on Schmidt's law (Schmidt
\cite{sch59}). The number $n\left( t \right)$ of stars that are formed per
time-step (of duration $10^6$ yr) is determined by the product
\begin{equation}
\label{sfn}
n\left(t\right) = \frac{\nu}{N_{tot}} \cdot \sum_{i=1}^{N_{tot}}
\rho_i^{\alpha} \left( t \right),
\end{equation}
where $\rho_i\left( t \right)$ is the local ISM density at time $t$ in a cell
of volume (50 pc)$^3$ and the sum goes over all cells in the volume
($N_{tot}$). The parameter $\nu$ determines the star formation efficiency,
and possible values for $\alpha$ range from 1 (SF proportional to gas
density) to 2 (SF triggered by cloud-cloud collisions, Larson
\cite{la91}). The factor $N_{tot}^{-1}$ was introduced to keep the number of
newly formed stars independent of the model resolution, so that Eq.~\ref{sfn}
gives $n\left( t \right) = \nu \left<\rho\left( t
\right)\right>^{\alpha}$ if the ISM is homogeneously distributed
(i.e. $\rho_i\left( t \right) = \left<\rho\left( t \right)\right>$ for all
$i$).

The mass of a newly formed star is chosen randomly, subject to the condition
that the mass distribution of all stars follows a Salpeter initial mass
function ($dN/dm \propto m^{-2.35}$) with lower and upper mass limits of
$m_{lo} = 0.1 \, \mathrm{M}_{\sun}$ and $m_{up} = 50 \, \mathrm{M}_{\sun}$,
respectively. The average mass $\left<m\right>$ of a star is given by the
initial mass function (IMF) as
\begin{equation}
\left<m\right> = \frac{\int_{m_{lo}}^{m_{up}} m \cdot
dN}{\int_{m_{lo}}^{m_{up}} dN}.
\end{equation}
The number $n\left(t\right)$ of star formation events translates into an
average star formation rate (SFR) $\left<\psi\left(t\right)\right>$ at each
time-step by multiplying the average stellar mass $\left<m\right>$, i.e.
\begin{equation}
\label{avgsfr}
\left<\psi\left(t\right)\right> = n\left(t\right) \cdot \left<m\right> =
\frac{\nu}{N_{tot}} \cdot \sum_{i=1}^{N_{tot}} \rho_i^{\alpha} \left( t
\right) \cdot \frac{\int_{m_{lo}}^{m_{up}} m \cdot dN}{\int_{m_{lo}}^{m_{up}}
dN}.
\end{equation}

The cells that undergo star formation are also chosen randomly, though the
probability for a cell to get selected scales with its density. In this
prescription, patches of denser material, e.g. in the neighbourhood of a
supernova remnant, are predominantly chosen for star formation events. Note
that a cell is allowed to form stars only if it contains at least $50 \,
\mathrm{M}_{\sun}$ of gas. This restriction is imposed, so that SF is not
biased towards low mass stars. With the imposed limit, a star formed in a cell
can be of any mass in the range $0.1 - 50 \, \mathrm{M}_{\sun}$. Consequently,
no SF will occur until enough material has fallen into the volume to exceed
this limit. Due to the stochastic nature of our SF prescription, small
deviations from the average SFR in Eq.~\ref{avgsfr} have to be expected at
each time-step. Newborn stars inherit the abundance pattern of the ISM out
of which they formed, thus carrying information about the chemical
composition of the ISM at the place and time of their birth.

Low mass stars ($\le 1 \, \mathrm{M}_{\sun}$) do not evolve significantly
during the considered time but serve to lock up part of the total mass, thus
affecting the abundances of elements with respect to hydrogen. Stars
of intermediate mass ($\sim 1 - 10 \, \mathrm{M}_{\sun}$) return most of their
mass at the end of their stellar lifetime, leaving a white dwarf as stellar
remnant. Stars in the mass range of $10-50 \, \mathrm{M}_{\sun}$ are assumed
to explode as SNe~II, polluting the neighbouring ISM with their highly
enriched ejecta.  Stellar lifetimes are taken from the \emph{Geneva Stellar
Evolution and Nucleosynthesis Group} (cf. Schaller et al. \cite{sl92};
Schaerer et al. \cite{sr93a}; Schaerer et al. \cite{sr93b}; Charbonnel et
al. \cite{cb93}). Stellar yields of O, Mg, Si, Ca and Fe are taken from
Thielemann et al. (\cite{th96}) and Nomoto et al. (\cite{no97}) and are scaled
according to Samland~(\cite{sa98}) to account for the global chemical
enrichment of the Galaxy. In particular, Fe yields are reduced by a factor of
two.

SN events pollute the neighbouring ISM with their nucleosynthesis products and
sweep up the material in a spherical, chemically well mixed shell. Here, it is
assumed that each SN pollutes $\approx 5 \times 10^4 \, \mathrm{M}_{\sun}$ of
ISM (Ryan et al. \cite{ry96}; Shigeyama \& Tsujimoto \cite{sh98}). Stars which
form out of material enriched by a \emph{single} SN~II show an element
abundance pattern that is characteristic of the yields for the particular
progenitor of this SN~II. This will lead to a large scatter in element
abundance ratios ([el/Fe]) as long as local inhomogeneities caused by SN~II
events dominate the halo ISM. As time progresses, supernova remnants overlap
and the abundance pattern in each cell approaches the average defined by SN~II
yield patterns for different progenitor masses and the IMF. This leads to a
decrease of the scatter in element abundance ratios at later times.

To determine which intermediate mass stars form Type~Ia SN events, we adopt
the following simple procedure: With probability $P_{\mathrm{SN Ia}}$ a newly
formed intermediate mass star has a companion in the same mass range and one
of these two stars will end its life as SN~Ia. The mass of the companion
(again in the range $1-10 \, \mathrm{M}_{\sun}$ and following a Salpeter IMF)
is determined randomly and the SN~Ia event occurs after the less massive of
the two stars enters the red giant phase. Although this procedure admittedly
is rather simple, it has the advantage that the SN~Ia frequency is determined
by only one free parameter, namely the probability $P_{\mathrm{SN Ia}}$. This
parameter is chosen in such a way that the slope observed in [$\alpha$/Fe]
abundance ratios at [Fe/H] $\ge -1$ is reproduced. We chose $10 \,
\mathrm{M}_{\sun}$ as upper mass limit for SN~Ia progenitors to be consistent
with the lower mass limit of SN~II. Since we use such a simplified description
to determine the occurrence of SN~Ia, the upper mass limit of SN~Ia
progenitors is not very relevant: Assuming a Salpeter IMF, more than 99\% of
the binary systems will consist of stars with masses $\le 8 \,
\mathrm{M}_{\sun}$ each. Finally, the yields of Type~Ia SNe are taken from
Iwamoto et al. (\cite{iwetal99}, Model CDD2).  Note, that if revised,
i.e. reduced electron capture rates are applied (see Brachwitz et
al. \cite{brach00}), the yields of model CDD2 correspond to those of model
WDD2, which are more accurate in terms of central conditions.

\begin{table}
 \caption{Parameter values of the standard model.}
 \begin{tabular}[]{llr} \hline
   Parameter \rule[-2mm]{0mm}{6mm} & Description & Value \\
  \hline\rule[0mm]{0mm}{4mm} 
  $V$                  & modelled volume          & 8 kpc$^3$ \\
  $M_{tot}$            & total system mass        & $10^8\,\mathrm{M}_{\sun}$\\
  $\tau$               & infall decline timescale & $5$ Gyr \\
  $t_{max}$            & time of maximal infall   & $2$ Gyr \\
  $t_{end}$            & age of the system        & $14$ Gyr \\
  $\nu$                & SF efficiency            & $15$ \\
  $\alpha$             & exponent of SF law       & $1.5$ \\
  $M_{\mathrm{lo}}$    & lower IMF mass limit     & $0.1\,\mathrm{M}_{\sun}$ \\
  $M_{\mathrm{up}}$    & upper IMF mass limit     & $50\,\mathrm{M}_{\sun}$ \\
  $P_{\mathrm{SN Ia}}$ & SN~Ia probability        & $6 \cdot 10^{-3}$ \\
  \hline
 \end{tabular}
 \label{param}
\end{table}

In Table~\ref{param} we list the parameter values adopted for our standard
model.

\subsection{Treatment of r-process sources in the model}
\label{rssources}

An ideal element to trace the r-process enrichment of the ISM is the pure
r-process element Europium; approximately 97\% of the solar Eu abundance was
produced in r-process events (Burris et al.~\cite{br00}). Unfortunately, only
a small sample of Eu abundances at very low metallicities (i.e. [Fe/H] $\le
-2.5$) are available to date. In order to trace the r-process enrichment at
lower metallicities, the well studied element Barium is also used in this
investigation. Ba abundances in stars are dominated by the s-process, and only
$\approx 15\%$ of the solar Ba abundance was formed by rapid neutron capture
(Burris et al.~\cite{br00}). However, the r-process fraction [Ba$^r$/Fe] of Ba
abundances in stars can be easily calculated by subtracting the s-process
component from its total Ba abundance (c.f. Burris et al. \cite{br00}).  Since
we did not include s-process sources in our model, displayed Ba$^r$ abundances
in Sect.~\ref{enrichment} of model and halo stars only show the r-process
contribution to the total Ba abundances.

In the following, r-process yields of Eu and Ba are estimated under the
assumption of a robust r-process for nuclei more massive than Ba, i.e. only
one source is responsible for the enrichment of the ISM with r-process
elements beyond Z=56. In this work we focus only on these heavier neutron
capture elements, since the enrichment of lighter neutron capture elements may
require a second, independent source (e.g. Wasserburg et al.~\cite{wa96}).
Thus, r-process matter ejected in an r-process event is considered to consist
of pure r-process nuclei more massive than Ba. First, the mass fractions of Ba
and Eu are calculated with the help of Table~5 in Burris et al. (\cite{br00})
and the standard solar system element abundances of Anders \& Grevesse
(\cite{ag89}):
\begin{equation}
f_{\mathrm{Ba}} = \frac{N^r_{\mathrm{Ba}} \cdot <m_{\mathrm{Ba}}>}{\sum_i
N^r_i \cdot <m_i>} \approx 7.7 \cdot 10^{-2},
\end{equation}
\begin{equation}
f_{\mathrm{Eu}} = \frac{N^r_{\mathrm{Eu}} \cdot <m_{\mathrm{Eu}}>}{\sum_i
N^r_i \cdot <m_i>} \approx 1.2 \cdot 10^{-2},
\end{equation}
where $N^r_i$ is the number fraction of r-process nuclei of a nuclear species
$i$, $<m_i>$ its mean atomic weight and the sum goes over all elements beyond
Ba ($Z \ge 56$). Final yields are estimated by multiplying these mass
fractions with the mass $M_r$ ejected in an r-process event, i.e.
\begin{equation}
\label{yba}
Y_{\mathrm{Ba}} \approx 7.7 \cdot 10^{-2} \cdot M_r,
\end{equation}
\begin{equation}
\label{yeu}
Y_{\mathrm{Eu}} \approx 1.2 \cdot 10^{-2} \cdot M_r.
\end{equation}
The total ejected r-process matter, $M_r$, is either deduced from
observational evidence and/or theoretical considerations or is treated as a
free parameter that is adjusted so that the results of the chemical evolution
model are consistent with observations.

We now shortly discuss the implementation of lower-mass SNe ($8-10 \,
\mathrm{M}_{\sun}$), higher-mass SNe ($\ge 20 \, \mathrm{M}_{\sun}$) and NSM
as r-process sites in our model. We did not consider a hybrid model, where
different r-process sites contribute to the enrichment of the ISM with neutron
capture elements. Instead, it is assumed that for each of the three cases the
corresponding site is responsible for the whole r-process element inventory
(with $Z \ge 56$) of the Galaxy, i.e. is the major source of r-process
elements beyond Ba.

\subsubsection{r-process yields from core-collapse SNe in the mass range 
$8-10 \, \mathrm{M}_{\sun}$}
\label{sn8.model}

Lower mass core-collapse SNe ($8-10 \, \mathrm{M}_{\sun}$) are suspected to be
major r-process sources without contributing much iron-peak elements to the
enrichment of the ISM (Wheeler et al. \cite{wh98}). Recently, Sumiyoshi et
al. (\cite{sm01}) presented results of r-process nucleosynthesis occurring
during the artificially induced \emph{prompt} explosion of an $11 \,
\mathrm{M}_{\sun}$ star. The resulting distribution of r-process yields is in
reasonable agreement with solar r-process abundances for the heavier elements,
whereas lighter r-process elements (A $<100$) are underproduced. Since
elements in this mass-regime are often overproduced in neutrino driven wind
models, the authors speculate that lower-mass SNe~II might be responsible for
the production of massive r-process nuclei, whereas higher-mass SNe~II account
for the less massive r-process nuclei. Note however, that it is still doubtful
if the prompt explosion of a massive star may occur in reality (Liebend\"orfer
et al. \cite{li01}; Sumiyoshi et al. \cite{sm01}). Chemical evolution models
including lower mass SNe as r-process source were presented by Ishimaru \&
Wanajo (\cite{is99}) and Travaglio et al. (\cite{tr99}).

In the treatment of such lower-mass SNe, we proceed similar to Ishimaru \&
Wanajo (\cite{is99}): In model SN810, r-process nucleosynthesis is assumed to
occur in the mass range $8-10 \, \mathrm{M}_{\sun}$ with constant Ba and Eu
yields over the whole mass range and assuming a Salpeter IMF. The yields then
are deduced from the average [Eu/Fe] and [Ba$^r$/Fe] ratios (both $\approx
0.5$) of metal-poor halo stars: $Y_{\mathrm{Ba}} = 5.3 \cdot 10^{-7} \,
\mathrm{M}_{\sun}$ and $Y_{\mathrm{Eu}} = 8.3 \cdot 10^{-8} \,
\mathrm{M}_{\sun}$, which equals to a total of $\approx 6.9 \cdot 10^{-6} \,
\mathrm{M}_{\sun}$ of ejected r-process matter per event (Eqs.~\ref{yba} and
\ref{yeu}). Furthermore, we assume that the amount of $\alpha$- and iron peak
elements synthesized in these SN~II events are negligible (Hillebrandt et
al. \cite{hi84}).

\subsubsection{r-process yields from core-collapse SNe more massive than 
$20 \, \mathrm{M}_{\sun}$}
\label{sn25.model}

Nucleosynthesis of r-process elements might also occur in neutrino driven
winds or jets from nascent neutron-stars during the delayed explosion of
high-mass stars with masses $\ge 20 \, \mathrm{M}_{\sun}$ (e.g. Woosley \&
Hoffman \cite{wo92}; Takahashi et al. \cite{tk94}; Woosley et al. \cite{wo94};
Qian \& Woosley \cite{qi96}; Thompson et al. \cite{tp01}; Nagataki
\cite{na01}; Wanajo et al. \cite{wn01}, \cite{wn02}; Terasawa et al. 
\cite{te02}). The physical conditions needed for a robust r-process in
neutrino driven winds (such as high entropies, low electron fraction and/or
short dynamical timescales) are hard to achieve in present models, and lighter
r-process nuclei are often overproduced (e.g. Thompson et al. \cite{tp01}). As
an example, Wanajo et al. (\cite{wn02}) require a very massive and compact
proto-neutron star of $2.0 \, \mathrm{M}_{\sun}$ to reproduce the solar system
r-process abundance pattern of heavier neutron capture elements. By allowing
for strong magnetic fields, Thompson (\cite{tp03}) finds that the required
physical conditions are much more easily achieved than in previous
models. Yet, there exist still some considerable theoretical uncertainties and
only very high-mass stars seem to be capable of providing the environment for
a robust r-process which reproduces the abundance pattern of neutron-capture
elements beyond Ba.

Williams (\cite{wi87}) reported the detection of Ba absorption lines in the
spectra of SN~1987A (progenitor mass $\approx 20 \, \mathrm{M}_{\sun}$), and
Mazzali et al. (\cite{mz92}) noted that Ba was lacking at the very surface of
the ejecta. Because of this remarkable feature, Tsujimoto \& Shigeyama
(\cite{ts01}) identify SN~1987A as r-process site and estimate that $\approx 6
\cdot 10^{-6} \, \mathrm{M}_{\sun}$ of Ba were synthesized during the SN
event. However, the poor knowledge of the far UV radiation field in the
envelope of SN~1987A makes it difficult to determine a truly reliable Ba
abundance in its ejecta (Utrobin \& Chugai \cite{ut02}) and it is not yet
established without doubt that the detected Ba was formed in the SN
event. Complementary to the observation of Ba in SN~1987A, Tsujimoto et
al. (\cite{ts00}) deduce Ba and Eu yields from observations of metal-poor halo
stars and inhomogeneous chemical evolution models, suggesting that
core-collapse SNe in the mass range $20 -25 \, \mathrm{M}_{\sun}$ dominate the
production of r-process elements. They propose a Ba yield of $8.5 \cdot
10^{-6} \, \mathrm{M}_{\sun}$ for a $20 \, \mathrm{M}_{\sun}$ and $4.5 \cdot
10^{-8} \, \mathrm{M}_{\sun}$ for a $25 \, \mathrm{M}_{\sun}$ progenitor and a
Eu yield of $1.3 \cdot 10^{-6} \, \mathrm{M}_{\sun}$ and $7.0 \cdot 10^{-9} \,
\mathrm{M}_{\sun}$ for a $20$ and $25 \, \mathrm{M}_{\sun}$ progenitor,
respectively.

In a first model (SN2025), we adopt yields similar to the ones given by
Tsujimoto et al. (\cite{ts00}) and assume that SNe outside the indicated mass
range (i.e. $20-25 \, \mathrm{M}_{\sun}$) do not contribute significantly to
r-process nucleosynthesis. In this model, a $20 \, \mathrm{M}_{\sun}$ star
produces $4.3 \cdot 10^{-6} \, \mathrm{M}_{\sun}$ of Ba and $6.5 \cdot 10^{-7}
\, \mathrm{M}_{\sun}$ of Eu, whereas a $25 \, \mathrm{M}_{\sun}$ star yields
$2.3 \cdot 10^{-8} \, \mathrm{M}_{\sun}$ of Ba and $3.5 \cdot 10^{-9} \,
\mathrm{M}_{\sun}$ of Eu. For stars between $20$ and $25 \,
\mathrm{M}_{\sun}$, the yields are interpolated linearly. Thus, a SN~II in
this mass range ejects on average $\approx 3 \cdot 10^{-5} \,
\mathrm{M}_{\sun}$ of r-process matter. Compared to the r-process yields
proposed by Tsujimoto et al. (\cite{ts00}) our yields are reduced by a factor
of two. This is the case, since we had to scale the Fe yields of Thielemann et
al. (\cite{th96}) by the same factor to account for the global chemical
enrichment of the Galaxy.

As an alternative to model SN2025, we also calculated the chemical evolution
of the ISM with r-process yields from SNe~II in the mass range $20-50 \,
\mathrm{M}_{\sun}$ (model SN2050). The yields were chosen in such a way that
the whole range of r-process abundances in metal-poor halo stars are
reproduced in the model. The r-process yield of a $20 \, \mathrm{M}_{\sun}$
star was set to $1.5 \cdot 10^{-4}$, which is about three times the r-process
yield of a $20 \, \mathrm{M}_{\sun}$ star in model SN2025. This large amount
of ejected r-process matter requires a rapid decline in the r-process yields
of progenitors with masses $20 \, \mathrm{M}_{\sun} < m < 28 \,
\mathrm{M}_{\sun}$. Otherwise, the average [r/Fe] ratios of model stars do not
reproduce the observed average abundances of metal-poor halo stars ([r/Fe]
$\approx 0.4-0.5$). Yields of progenitors in the mass range $28-50 \,
\mathrm{M}_{\sun}$ then are assumed to be constant. On average, $1.2 \cdot
10^{-5} \, \mathrm{M}_{\sun}$ of r-process matter are ejected in each event.

\begin{table}
 \caption{Adopted Ba ($Y_{\mathrm{Ba}}\left(m\right)$) and Eu
          ($Y_{\mathrm{Eu}}\left(m\right)$) yields and ejected r-process
          matter ($M_r$) as function of progenitor mass $m$ of models SN810,
          SN2025 and SN2050. Yields in the mass range $8-10 \,
          \mathrm{M}_{\sun}$ are assumed to be constant (SN810). Yields in the
          mass range $20-25 \, \mathrm{M}_{\sun}$ are linearly interpolated
          (SN2025). Yields of model SN2050 exhibit a more complex behaviour: A
          rapid decline in the mass range $20-28 \, \mathrm{M}_{\sun}$ is
          assumed. Progenitors with masses $28-50 \, \mathrm{M}_{\sun}$ all
          have the same, constant yield.}
 \begin{tabular}[]{lccc} \hline
     $m$ & 
     $Y_{\mathrm{Ba}}\left(m\right) \, \left[\mathrm{M}_{\sun}\right]$ &
     $Y_{\mathrm{Eu}}\left(m\right) \, \left[\mathrm{M}_{\sun}\right]$ &
     $M_r \, \left[\mathrm{M}_{\sun}\right]$ \rule[-2mm]{0mm}{6mm} \\
  \hline
      8$^a$ & $5.3 \cdot 10^{-7}$ & $8.3 \cdot 10^{-8}$ & $6.9 \cdot 10^{-6}$ \\
     10 & $5.3 \cdot 10^{-7}$ & $8.3 \cdot 10^{-8}$ & $6.9 \cdot 10^{-6}$ \\
  \hline
     20$^b$ & $4.3 \cdot 10^{-6}$ & $6.5 \cdot 10^{-7}$ & $5.4 \cdot 10^{-5}$ \\
     25 & $2.3 \cdot 10^{-8}$ & $3.5 \cdot 10^{-9}$ & $2.9 \cdot 10^{-7}$ \\
  \hline
     20$^c$ & $1.1 \cdot 10^{-5}$ & $1.8 \cdot 10^{-6}$ & $1.5 \cdot 10^{-4}$ \\
     21 & $3.3 \cdot 10^{-6}$ & $5.1 \cdot 10^{-7}$ & $4.2 \cdot 10^{-5}$ \\
     22 & $7.9 \cdot 10^{-7}$ & $1.2 \cdot 10^{-7}$ & $1.0 \cdot 10^{-5}$ \\
     23 & $2.5 \cdot 10^{-7}$ & $3.8 \cdot 10^{-8}$ & $3.2 \cdot 10^{-6}$ \\
     24 & $9.2 \cdot 10^{-8}$ & $1.4 \cdot 10^{-8}$ & $1.2 \cdot 10^{-6}$ \\
     25 & $4.8 \cdot 10^{-8}$ & $7.5 \cdot 10^{-9}$ & $6.2 \cdot 10^{-7}$ \\
     26 & $2.7 \cdot 10^{-8}$ & $4.3 \cdot 10^{-9}$ & $3.6 \cdot 10^{-7}$ \\
     27 & $1.8 \cdot 10^{-8}$ & $2.7 \cdot 10^{-9}$ & $2.3 \cdot 10^{-7}$ \\
     28 & $1.1 \cdot 10^{-8}$ & $1.7 \cdot 10^{-9}$ & $1.4 \cdot 10^{-7}$ \\
     50 & $1.1 \cdot 10^{-8}$ & $1.7 \cdot 10^{-9}$ & $1.4 \cdot 10^{-7}$ \\
  \hline
 \end{tabular}
 Remarks: $^a$Model SN810, $^b$Model SN2025, $^c$Model SN2050
 \label{snparam}
\end{table}

Table~\ref{snparam} lists our Ba yields $Y_{\mathrm{Ba}}\left(m\right)$ and Eu
yields $Y_{\mathrm{Eu}}\left(m\right)$ and total ejected r-process matter
$M_r$ as function of progenitor mass $m$ of models SN810, SN2025 and SN2050.
We point out that the yields adopted in this work are chosen in such a way
that observational constraints from metal-poor halo stars are respected.
Whether these yields are feasible would have to be tested with the help of
self-consistent core-collapse SN models, which, unfortunately, are not
available to date.

\subsubsection{r-process yields from neutron star mergers}
\label{nsm.model}

Another major source of r-process elements might be neutron star mergers
(Freiburghaus et al. \cite{fr99b}; Rosswog et al. \cite{ro99}; \cite{ro00}). A
scheme similar to the determination of SN~Ia events was applied for the
determination of their occurrence (c.f. Sect.~\ref{ice}): With probability
$P_{\mathrm{NSM}}$ a newly formed high mass star (in the range $10 - 50 \,
\mathrm{M}_{\sun}$) has a companion in the same mass range. It is assumed
that, subsequent to both SN~II events, the remaining neutron stars will
eventually merge. The time $t_c$ that is needed for the coalescence of the two
neutron stars and the probability $P_{\mathrm{NSM}}$ for the occurrence of NSM
events are treated as free parameters which can be adjusted within given
observational estimates on the coalescence rate of binary compact objects and
merger timescales (van den Heuvel \& Lorimer \cite{vdH96}; Kalogera \& Lorimer
\cite{ka00}; Belczynski et al. \cite{bl02}). As was the case for SN~Ia events,
this treatment is very simple, but has the advantage that all difficulties
associated with the determination of formation channels of neutron star
binaries (or neutron star / black hole binaries) and the corresponding
formation rates are combined in only one free parameter.  Furthermore, the
lower mass boundary for neutron star formation (here $10 \,
\mathrm{M}_{\sun}$) is of little importance: The decisive term for the
enrichment of the ISM with r-process material in our model is the NSM
rate. Since a decrease of the lower mass boundary results in an increase of
the NSM rate, a simple adjustment of the parameter $P_{\mathrm{NSM}}$ is
sufficient to retrieve a NSM rate consistent with constraints on the galactic
rate. In the model the ejected r-process matter is also mixed with $\approx 5
\cdot 10^4 \, \mathrm{M}_{\sun}$ of ISM and is assumed to be distributed in a
spherical, chemically well mixed shell. This might not be true, however, since
it is well conceivable that r-process matter is ejected only in the orbital
plane of the merging neutron stars. In this case, the ejecta would be
distributed over a smaller volume and would consequently mix with a smaller
amount of ISM than the one assumed in our model. As will be seen below
(Sect.~\ref{nsmdom}), this would lead to even higher [r/Fe] ratios in the
computed model stars. Thus, our assumptions are in favour of the NSM scenario
and changing the mixing mechanism in our model would only aggravate the
problems associated with the high r-process yields from NSM.

Coalescence timescales for neutron star mergers are typically estimated to be
of the order $100 - 1000$ Myr (e.g. Portegies~Zwart \& Yungel'son \cite{pr98};
Fryer et al. \cite{fy99}). Recently, Belczynski et al. (\cite{bl02}) suggested
a \emph{dominating} population of short lived neutron star binaries with
merger times less than 1 Myr. This population of neutron star mergers might be
formed through channels involving mass-transfer episodes from helium stars,
leading to tightly bound binary systems with very short orbit decay
timescales.  However, this result depends on the occurrence of a common
envelope (CE) phase of the progenitor He stars, which was treated in a
simplified manner by Belczynski et al. (\cite{bl02}). The authors caution
that detailed hydrodynamical calculations of the CE phase still have to
confirm these results. In view of these uncertainties, we adopt two
different coalescence timescales $t_c$, namely 1 and 100 Myr. In each
case, it is assumed that \emph{all} neutron star binaries merge within this
time. Coalescence timescales of the order 1 Gyr have not been considered
since they are not consistent with observations of neutron capture elements in
ultra metal-poor stars (c.f. Sect~\ref{nsmdom}).

\begin{table}
 \caption{Constraints on the Galactic NSM rate $f_{\mathrm{NSM}}$ and ejected 
          r-process matter $M_r$.}
 \begin{tabular}[]{lc} \hline
     Reference & $f_{\mathrm{NSM}} \, [\mathrm{yr}^{-1}]$ \rule[-2mm]{0mm}{6mm} \\
  \hline
     Heuvel \& Lorimer (\cite{vdH96})  & $8 \cdot 10^{-6}$ \rule[0mm]{0mm}{4mm}\\
     Kalogera \& Lorimer (\cite{ka00}) & $\le \left( 7.5 \cdot 10^{-7} - 1.5 \cdot 10^{-5} \right)$ \\
     Belczynski et al. (\cite{bl02})   & $10^{-6} - 3 \cdot 10^{-4}$  \\
  \hline
     Reference & $M_r$ $[\, \mathrm{M}_{\sun}]$ 
     \rule[-2mm]{0mm}{6mm} \\
  \hline
     Rosswog et al. (\cite{ro99}) & $10^{-3} - 10^{-2}$\rule[0mm]{0mm}{4mm}\\
     Oechslin et al. (\cite{oe02}) & $\ge \left( 5 \cdot 10^{-5} - 2 \cdot 10^{-4} \right)$\\
  \hline
 \end{tabular}
 \label{nsmcons}
\end{table}

Constraints on the Galactic NSM rate are also controversial. Van den Heuvel \&
Lorimer (\cite{vdH96}) estimate a rate of approximately $8 \cdot 10^{-6} \,
\mathrm{yr}^{-1}$, Kalogera \& Lorimer (\cite{ka00}) give an upper limit of
$\left(7.5 \cdot 10^{-7} - 1.5 \cdot 10^{-5}\right) \, \mathrm{yr}^{-1}$
whereas Belczynski et al. (\cite{bl02}) get rates in the range $\left(10^{-6}
- 3 \cdot 10^{-4}\right) \, \mathrm{yr}^{-1}$. In Newtonian calculations, the
amount $M_r$ of r-process matter ejected in a NSM event is of the order of a
few times $10^{-3} - 10^{-2} \, \mathrm{M}_{\sun}$, depending on the initial 
configuration of the binary system (Rosswog et al. \cite{ro99}). Taking 
general relativistic effects into account, Oechslin et al. (\cite{oe02}) find
$5 \cdot 10^{-5} - 2 \cdot 10^{-4} \, \mathrm{M}_{\sun}$ as lower limit for
$M_r$. All constraints are again listed in Table~\ref{nsmcons}.

\begin{table}
 \caption{Parameter values adopted for NSM events (see text for details). For
          each value of $P_\mathrm{NSM}$, two models with the indicated
          coalescence timescales $t_c$ have been calculated.}
 \begin{tabular}[]{lccc} \hline
     $P_{\mathrm{NSM}}$ & $f_{\mathrm{NSM}} \, [\mathrm{yr}^{-1}]$ & 
     $M_r \, [\mathrm{M}_{\sun}]$ & $t_c$ [Myr] \rule[-2mm]{0mm}{6mm} \\
  \hline
     $4.1 \cdot 10^{-2}$ & $2 \cdot 10^{-3}$ & $10^{-4}$ & 1, 100 \\
     $4.0 \cdot 10^{-3}$ & $2 \cdot 10^{-4}$ & $10^{-3}$ & 1, 100 \\
     $4.0 \cdot 10^{-4}$ & $2 \cdot 10^{-5}$ & $10^{-2}$ & 1, 100 \\
     $3.9 \cdot 10^{-5}$ & $2 \cdot 10^{-6}$ & $10^{-1}$ & 1, 100 \\
  \hline
 \end{tabular}
 \label{nsmparam}
\end{table}

Table~\ref{nsmparam} lists the parameter values adopted for the NSM
probability $P_{\mathrm{NSM}}$, the resulting average NSM rate
$f_{\mathrm{NSM}}$, the amount of r-process matter $M_r$ ejected in each event
and coalescence timescales $t_c$. After a value for $P_{\mathrm{NSM}}$ has
been chosen, the coalescence timescale is the only free parameter remaining in
the model since the actual NSM rate is determined by the NSM probability and
the SF rate (resulting from the parameter values given in Table~\ref{param}).
In addition, the NSM rate and the amount of ejected r-process matter are
tightly correlated, since the total amount of r-process matter in the Galaxy
($\approx 10^4 \, \mathrm{M}_{\sun}$, Wallerstein et al. \cite{wl97}) has to
be reproduced. Thus, higher NSM rates require that less r-process matter is
ejected in each event, and vice versa.

Comparing the values in Table~\ref{nsmparam} with the constraints listed in
Table~\ref{nsmcons} reveals that the NSM rate of the first row in
Table~\ref{nsmparam} is too high by a factor of 10-100 and that the
corresponding low value of $10^{-4} \, \mathrm{M}_{\sun}$ of ejected r-process
matter is close to the lower limit given by Oechslin et al. (\cite{oe02}). NSM
rates listed in the three lower rows seem to be consistent with the galactic
NSM rate. However, the ejected r-process matter in the last row is clearly
beyond the upper limit of allowed values.


\section{Enrichment of the ISM with r-process elements}
\label{enrichment}

\subsection{SN~II as dominating r-process sites}

\begin{figure*}
 \resizebox{\hsize}{!}{\includegraphics{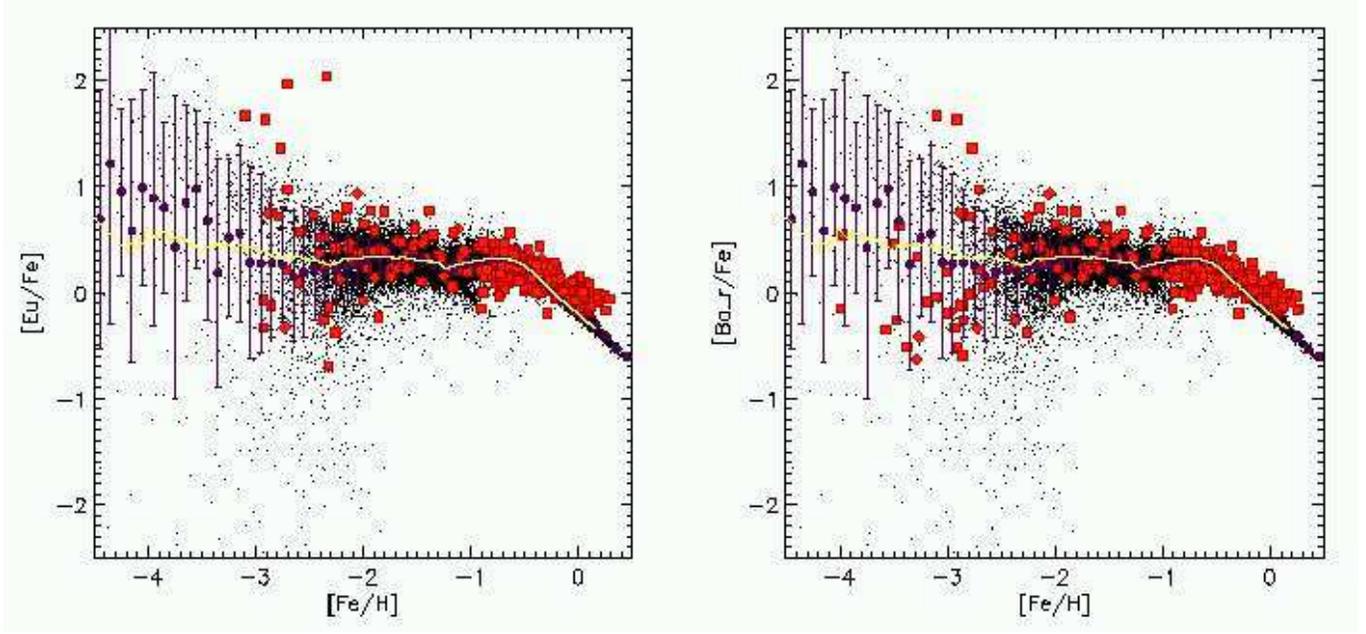}}
 \caption{Evolution of [Eu/Fe] and [Ba$^r$/Fe] abundances as function of
          metallicity [Fe/H]. Lower-mass SNe~II ($8 - 10 \,
          \mathrm{M}_{\sun}$, Model SN810) are assumed to be the dominating
          r-process sources. Black dots denote model stars, observations are
          marked by filled squares and diamonds (see text). Average ISM
          abundances are marked by a continuous line. Filled circles with
          error bars denote average abundances of model stars and their
          standard deviation in [Fe/H] bins with binsize 0.1 dex.}
 \label{sn810.fig}
\end{figure*}

\begin{figure*}
 \resizebox{\hsize}{!}{\includegraphics{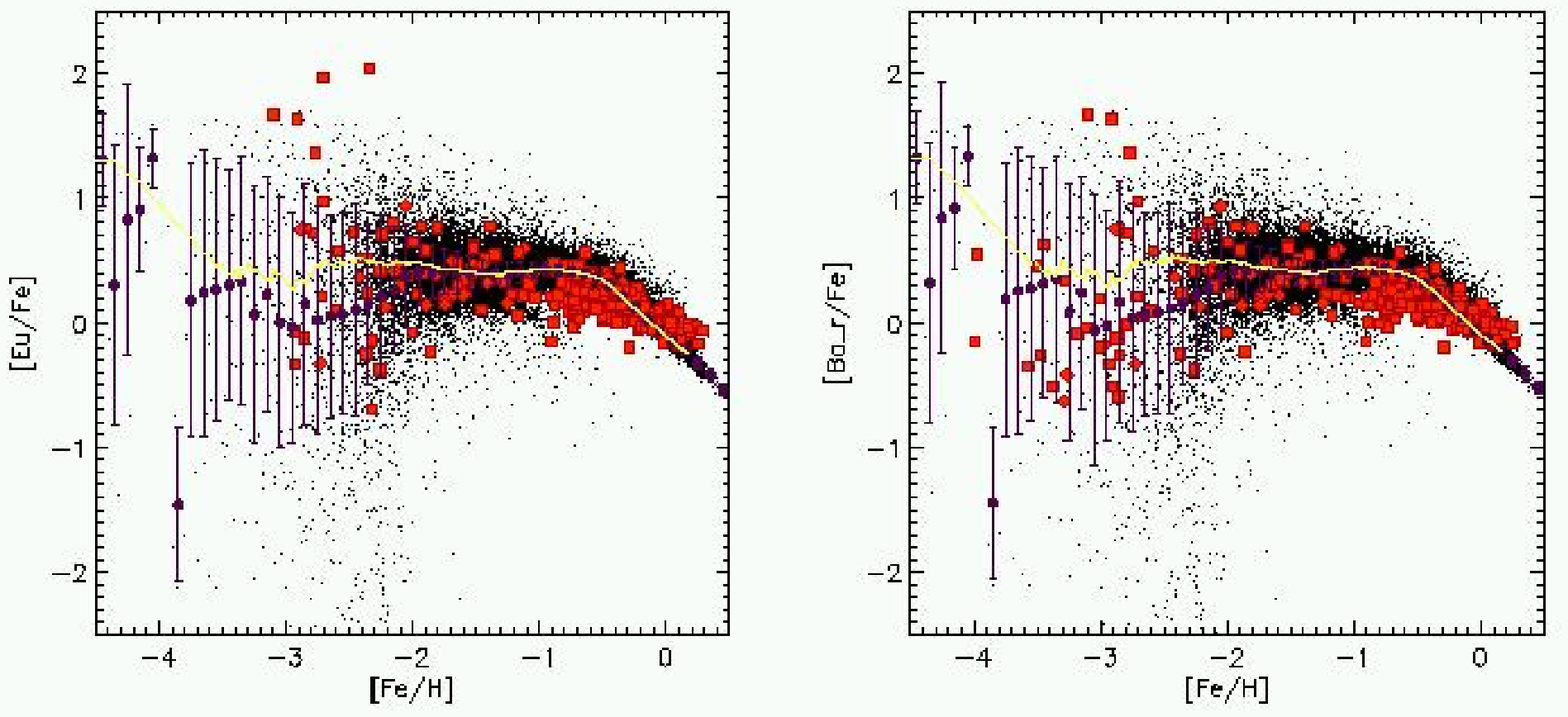}}
 \caption{Evolution of [Eu/Fe] and [Ba$^r$/Fe] abundances as function of
          metallicity [Fe/H]. Higher-mass SNe~II ($20 - 25 \,
          \mathrm{M}_{\sun}$, Model SN2025) are assumed to be the dominating
          r-process sources. Symbols are as in Fig.~\ref{sn810.fig}.}
 \label{sn2025.fig}
\end{figure*}

\begin{figure*}
 \resizebox{\hsize}{!}{\includegraphics{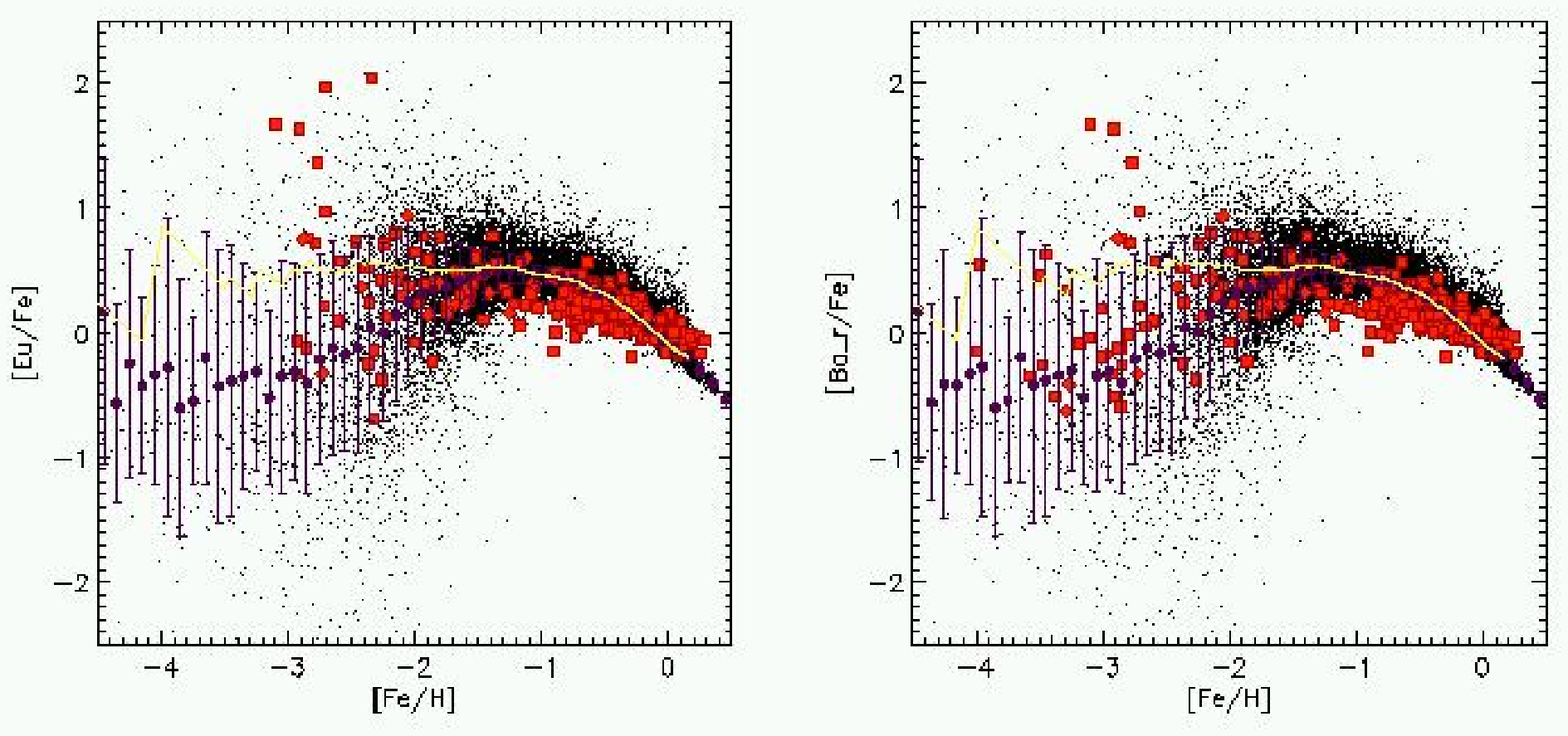}}
 \caption{Evolution of [Eu/Fe] and [Ba$^r$/Fe] abundances as function of
          metallicity [Fe/H]. Higher-mass SNe~II ($20 - 50 \,
          \mathrm{M}_{\sun}$, Model SN2050) are assumed to be the dominating
          r-process sources. Symbols are as in Fig.~\ref{sn810.fig}.}
 \label{sn2050.fig}
\end{figure*}

In this section, the enrichment of the ISM with neutron capture elements is
discussed under the assumption, that the dominating r-process sources are
either lower-mass SNe~II ($8 - 10 \, \mathrm{M}_{\sun}$) or higher-mass SNe~II
($> 20 \, \mathrm{M}_{\sun}$). The results of models SN810, SN2025 and SN2050
are shown in Figs.~\ref{sn810.fig}~--~\ref{sn2050.fig}, respectively. The
figures show the evolution of [Eu/Fe] and [Ba$^r$/Fe] as function of
metallicity [Fe/H]. Model stars are shown as black dots, whereas observations
are indicated by filled red squares and diamonds. Observations are taken from
Peterson et al. (\cite{pe90}), Gratton \& Sneden (\cite{gr91a}, \cite{gr91b}),
Ryan et al. (\cite{ry91}), Edvardsson et al. (\cite{ed93}), Fran\c{c}ois et
al. (\cite{fr93}), Beveridge \& Sneden (\cite{be94}), McWilliam et
al. (\cite{mw95a}), Ryan et al. (\cite{ry96}), Jehin et al. (\cite{je99}),
Aoki et al. (\cite{ao00}), Burris et al. (\cite{br00}), Mashonkina \& Gehren
(\cite{mg00}), Sneden et al. (\cite{sn00}), Mashonkina \& Gehren
(\cite{mg01}), Mishenina \& Kovtyukh (\cite{mi01}), Koch \& Edvardsson
(\cite{ke02}) and Stephens \& Boesgaard (\cite{st02}).  Single observations of
stars are marked by a square. In the case where multiple observations of a
star are present, we plot the most recent one if all observations were
published before the year 2000 (also marked by squares). If several more
recent observations are available, the given element abundances are averaged
(marked by diamonds). The yellow line shows the average element abundances in
the model ISM and can directly be compared to classical chemical evolution
models, which assume that the ISM is well mixed at all times. Purple filled
circles with error bars denote the average [r/Fe] ratios of model stars and
their standard deviation in [Fe/H] bins with binsize 0.1 dex. Note, that the
[Ba$^r$/Fe] plots only show the r-process contribution to the total Ba
abundances of halo stars. According to Burris et al. (\cite{br00}), r-process
Ba abundances can be computed by removing the s-process contribution to Ba in
stars with [Fe/H] $> -2.5$, if Eu abundances have also been determined. For
stars with metallicities [Fe/H] $< -2.5$, it can be assumed that the whole Ba
inventory is of pure r-process origin. Thus, published Ba abundances of such
metal-poor halo stars simply have to be renormalized to the level of the solar
r-process fraction (i.e. $\approx 15\%$ of the total solar Ba abundance). Care
has been taken to remove known carbon stars from our sample. Such stars mostly
show unusually large Ba abundances, which are thought to originate from mass
transfer of s-process enriched matter in binary systems.

The evolution of r-process elements shown in
Figs.~\ref{sn810.fig}~--~\ref{sn2050.fig} are qualitatively very similar. At
very low metallicities ([Fe/H] $\le -2.5$), a large scatter in abundances of
model stars is visible. This scatter is due to chemical inhomogeneities in the
early ISM (c.f. Paper~I). The scatter decreases as the mixing of the ISM
improves and finally reaches the IMF averaged mean. At this stage, the ISM can
be considered well mixed and the further evolution is comparable to the one of
classic chemical evolution models. In the following, we point out some
important features of the ISM enrichment resulting from models SN810, SN2025
and SN2050:

\begin{enumerate}
\item In all models, r-process elements appear very early in the enrichment of
the ISM. Some model stars with r-process abundances exist even at
metallicities [Fe/H] $\le -4$, which is in agreement with observed Ba
abundances in ultra metal-poor stars. Neutron capture elements at such low
metallicities are considered to be of pure r-process origin, since s-process
enrichment is expected to occur at a later stage in the chemical evolution of
the Galaxy (e.g. Travaglio et al. \cite{tr99}).
\item For metallicities [Fe/H] $\ge -2$, the scatter in [Eu/Fe] and
[Ba$^r$/Fe] abundances of model stars is comparable to the one seen in
observations. The main difference between the higher-mass (models SN2025 and
SN2050, Fig.~\ref{sn2025.fig} and \ref{sn2050.fig}) and lower-mass (model
SN810, Fig.~\ref{sn810.fig}) SNe~II scenarios is the larger scatter in [r/Fe]
at [Fe/H] $\approx -3$ and the slightly more pronounced scatter at higher
metallicities of models SN2025 and SN2050. One might argue, that the model
stars in Figs.~\ref{sn2025.fig} and \ref{sn2050.fig} at [Fe/H] $\ge -2$ with
high [r/Fe] abundances ($\ge 1$) should also be visible in the observational
data. The bulk of model stars, however, has abundances [r/Fe] $\approx
0.4-0.5$ and it would thus be very unlikely to observe stars with high [r/Fe]
ratios in this metallicity regime. The situation below [Fe/H] $= -2$ is
different: Only few model stars are present in this metallicity range and
their [r/Fe] abundances are evenly distributed and do not clump around the IMF
averaged value.
\item Models SN810 and SN2025 fail to reproduce the two metal-poor stars with
the highest Eu abundances ([Eu/Fe] $> +2.0$ at [Fe/H] $\approx
-2.5$). The existence of such ultra r-process enhanced stars might pose a
serious problem for the SN scenario, since an unusual large amount of ejected
r-process matter ($\ge 10^{-4} \, \mathrm{M}_{\sun}$) is required to reproduce
these observations in chemical evolution models (model SN2050,
Fig.~\ref{sn2050.fig}). In the neutrino driven wind model of Wanajo et al. 
(\cite{wn01}), massive proto-neutron stars with a high compaction ratio are
required to achieve r-process yields of this order of magnitude. However,
chemical evolution models constrain such large r-process yields to only a
small progenitor mass range (see Table~\ref{snparam}). Otherwise, the total
r-process inventory of the Galaxy is exceeded.

On the other hand, Qian \& Wasserburg (\cite{qiwa01}, \cite{qiwa03}) argue
that the highest Eu and Ba abundances in metal-poor halo stars may originate
from the dumping of r-process enriched matter from a high-mass companion onto
the surface of a low mass star. This explanation requires that during the
SN~II event of the high-mass companion, r-process elements but no iron-peak
nuclei are produced. Possible candidates for this kind of SN~II events are
progenitors in the mass range $8-10 \, \mathrm{M}_{\sun}$; see e.g. Qian \&
Wasserburg (\cite{qiwa03}) for a discussion of this mechanism. To date, our
model is not able to handle the occurrence of surface contamination in binary
systems, so that the results of model SN810 do not directly apply to this
scenario. We speculate however, that if this mechanism would be at work during
the evolution of the Galaxy, we might also expect stars with high [r/Fe]
abundances at metallicities [Fe/H] $> -2$. No such stars have been observed to
date. Yet, since many abundance studies primarily target metal-poor stars, it
is possible that r-process enriched, metal-rich stars simply have been
missed. Note, that if iron is produced in the SN~II event of the companion,
the resulting surface abundance will be a mixture of the intrinsic r-process
abundances and the yields of the SN~II. In this case, large r-process yields
are still required to enrich the companion to the high level seen in
observations.
\item The models also predicts some stars with very low [Eu/Fe] and [Ba$^r$/Fe]
ratios at [Fe/H] $\le -2$, which are not observed to date. However, if 
stars with [r/Fe] $\le -1$ exist in the galactic halo, their Eu and Ba lines
may be too weak to be detectable (c.f. observational limits quoted in
Travaglio et al. (\cite{tr01})). In the models, stars with very low [Eu/Fe]
ratios are inevitably produced if only a limited mass-range of SNe~II
contributes to the r-process enrichment. In this case, pockets without or with
only small amounts of r-process elements may form in the halo ISM if primarily
SNe~II without r-process yields contributed to the local enrichment.
\item Model stars in Fig.~\ref{sn810.fig} with high r-process abundances
([r/Fe] $\ge +2$) at metallicities below [Fe/H] $\le -3.5$ owe their origin to
the same phenomenon. Here, lower-mass SNe~II are the primary source of
r-process elements, without contributing much iron-peak nuclei to the
enrichment of the ISM. Thus, pockets with large r-process abundances may form
during the early enrichment of the ISM, similar to pockets with large Fe
abundances. However, [r/Fe] ratios soon converge to the average [r/Fe] ratios
of metal-poor halo stars and from then on give a good fit to observations. It
is interesting to note, that this behaviour is typical for inhomogeneous
chemical evolution models and is not present in classic chemical evolution
models which assume that the ISM is well mixed at all times (e.g. Travaglio et
al. \cite{tr99}). This simplifying assumption leads to a slow increase of the
average ISM [r/Fe] abundances, since r-process yields from $8-10 \,
\mathrm{M}_{\sun}$ are instantaneously mixed with iron yields from $10-50 \,
\mathrm{M}_{\sun}$.

We briefly note here, that the fit of model SN810 to observations of
metal-poor halo stars can be improved, if infall of pre-enriched ISM with
[Fe/H] $=-3$ is assumed (e.g. Qian \& Wasserburg \cite{qiwa02}). In this case,
model stars with high [r/Fe] abundances at [Fe/H] $\sim -4$ are shifted to
[Fe/H] $=-3$, where such r-process enriched stars are observed. This is the
case only for model SN810. The fit to observations of models SN2025, SN2050
and the NSM models (see Sect.~\ref{nsmdom}) are not changed significantly by
the assumption of metal-rich infall.
\item The evolution of average [r/Fe] abundances of model stars (purple
circles) as function of metallicity is an important tool to compare model
results with observations: Slopes in the distribution of model stars, which
might be hidden in the full [r/Fe] plots, will show up in the distribution of
these averaged abundances and may be compared to trends seen in observational
data. Note, that the evolution of average ISM abundances (yellow line) is
generally different from the evolution of average model star abundances. This
is the case, since ISM abundances are averaged over all cells in our volume,
which is equivalent to assuming a well mixed ISM at all times and thus may be
compared to classic chemical evolution models. Average model star abundances,
on the other hand, are still affected by chemical inhomogeneities, since they
depend on the \emph{number} of stars with a given [r/Fe] abundance.

This is demonstrated in Fig.~\ref{sn810.fig} (model SN810). Although the
calculated standard deviation of model stars is large (due to the low number
of model stars with metallicities below [Fe/H] $<-3$), the average model star
abundances rise with decreasing metallicity. Thus, model SN810 would predict a
rising slope in [Ba$^r$/Fe], contrary to the one present in observations of
metal-poor halo stars. In model SN2025, the average model abundances stay more
or less constant, whereas in model SN2050, average abundances give a nice fit to
the observational data. Note, that the empty bin at [Fe/H] $= -4$ in
Fig.~\ref{sn2025.fig} only contains one model star.
\end{enumerate}

In summary, core-collapse SNe seem to be a valid source of r-process elements
from the point of view of chemical evolution, since the enrichment of the ISM
in the cases discussed above is in qualitative, if not necessarily
quantitative, agreement with observations. Based on the evolution of average
model star abundances and the range of the scatter at [Fe/H] $\ge -3$, model
SN2050 gives the best fit to observations, whereas model SN810 gives the
worst. Note however, that the initial scatter in [r/Fe] ratios strongly
depends on the yields as function of progenitor mass. As was discussed in
Argast et al.  (\cite{ar02}), the progenitor mass dependence of stellar Fe
yields are not known to date and the distribution of observed element
abundances as function of metallicity can easily be reproduced by the
appropriate choice of Fe yields. Additionally, r-process yields in this work
are chosen completely \emph{ad hoc} and yet lack any theoretical backup. In
view of these uncertainties, we conclude in accordance with Ishimaru \& Wanajo
(\cite{is99}), that it is not possible to date to rule out either lower-mass
or higher-mass SNe~II within the framework of inhomogeneous chemical
evolution.

\subsection{NSM as dominating r-process sites}
\label{nsmdom}

\begin{figure*}
 \resizebox{\hsize}{!}{\includegraphics{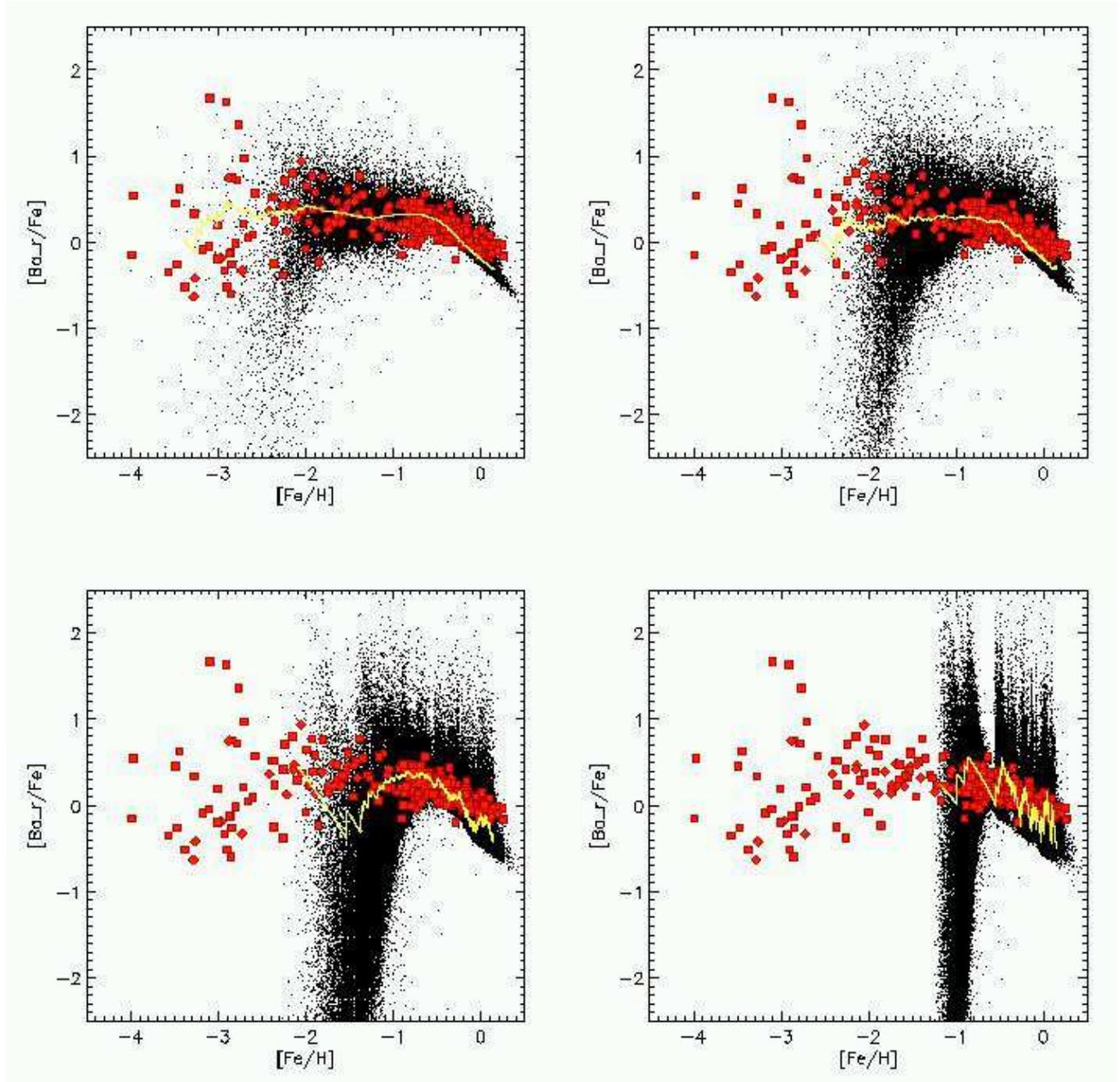}}
 \caption{[Ba$^r$/Fe] vs. [Fe/H] for the NSM rates $2 \cdot 10^{-3} \,
          \mathrm{yr}^{-1}$, $2 \cdot 10^{-4} \, \mathrm{yr}^{-1}$, $2 \cdot
          10^{-5} \, \mathrm{yr}^{-1}$ and $2 \cdot 10^{-6} \,
          \mathrm{yr}^{-1}$ (from left to right and top to bottom). The
          coalescence timescale adopted in these cases is $t_c=10^6$ yr.}
  \label{nsm001.fig}
\end{figure*}

\begin{figure*}
 \resizebox{\hsize}{!}{\includegraphics{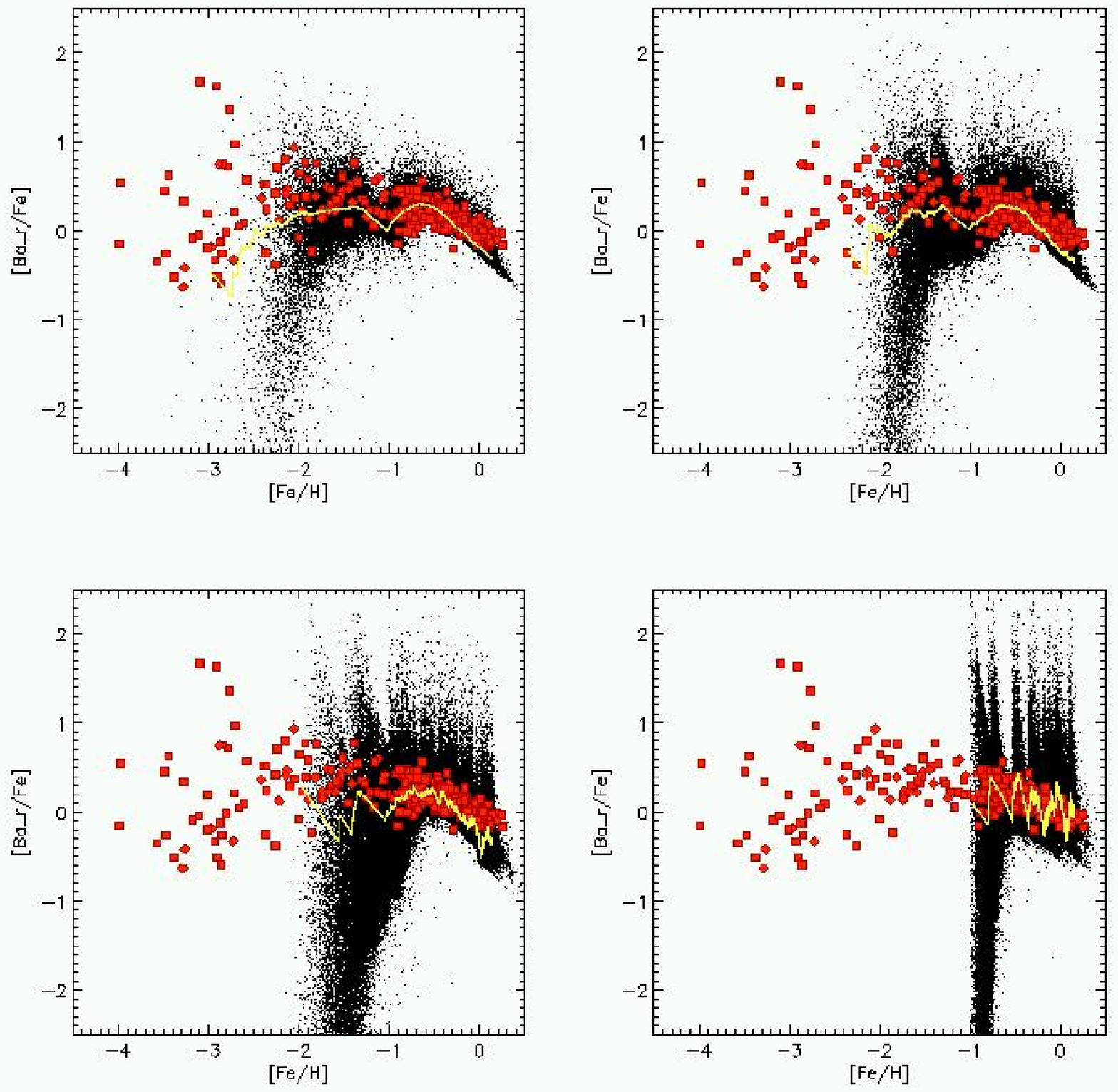}}
 \caption{Same as Fig.~\ref{nsm001.fig} but with $t_c = 10^8$ yr.}
 \label{nsm100.fig}
\end{figure*}

The enrichment of the ISM with r-process elements in the case of NSM acting as
major r-process sources is discussed in the following. As already mentioned in
Sect.~\ref{nsm.model}, eight model-runs were carried out in total, assuming
NSM rates ranging from $2 \cdot 10^{-3} \, \mathrm{yr}^{-1}$ to $2 \cdot
10^{-6} \, \mathrm{yr}^{-2}$ and coalescence time-scales of 1 and 100 Myr (see
Table~\ref{nsmparam}). A condensed overview of all models displaying the
evolution of [Ba$^r$/Fe] is shown in Figs.~\ref{nsm001.fig} and
\ref{nsm100.fig}.  The evolution of r-process abundances is strikingly
different from the case in which r-process nucleosynthesis occurs in SNe~II (see
Figs.\ref{sn810.fig}~--~\ref{sn2050.fig}). All models with NSM as r-process
sources fail to reproduce observations.

As a representative example, the model shown in the upper right panel of
Fig.~\ref{nsm001.fig} is discussed in the following. It was calculated with a
NSM rate of $2 \cdot 10^{-4} \, \mathrm{yr}^{-1}$, a total of $10^{-3} \,
\mathrm{M}_{\sun}$ of ejected r-process matter and a coalescence time-scale
$t_c$ of 1 Myr. Note, that the case discussed here assumes a NSM rate which is
at the upper limit given by present estimates of the galactic NSM rate:
Estimates range from $8 \cdot 10^{-6} \, \mathrm{yr}^{-1}$ (van den Heuvel \&
Lorimer \cite{vdH96}) to $\left(10^{-6} - 3 \cdot 10^{-4}\right) \,
\mathrm{yr}^{-1}$ (Belczynski et al. \cite{bl02}) and Kalogera \& Lorimer
(\cite{ka00}) even give an \emph{upper limit} of $\left(7.5 \cdot 10^{-7} -
1.5 \cdot 10^{-5}\right) \, \mathrm{yr}^{-1}$. Furthermore, the coalescence
timescale adopted in this model is only 1 Myr, which is two to three orders of
magnitude lower than the classical estimate of $100 - 1000$ Myr
(e.g. Portegies~Zwart \& Yungel'son \cite{pr98}; Fryer et
al. \cite{fy99}). However, Belczynski et al. (\cite{bl02}) suggest the
existence of a \emph{dominating} population of short lived neutron star
binaries with merger times less than 1 Myr.

\begin{figure*}
 \resizebox{\hsize}{!}{\includegraphics{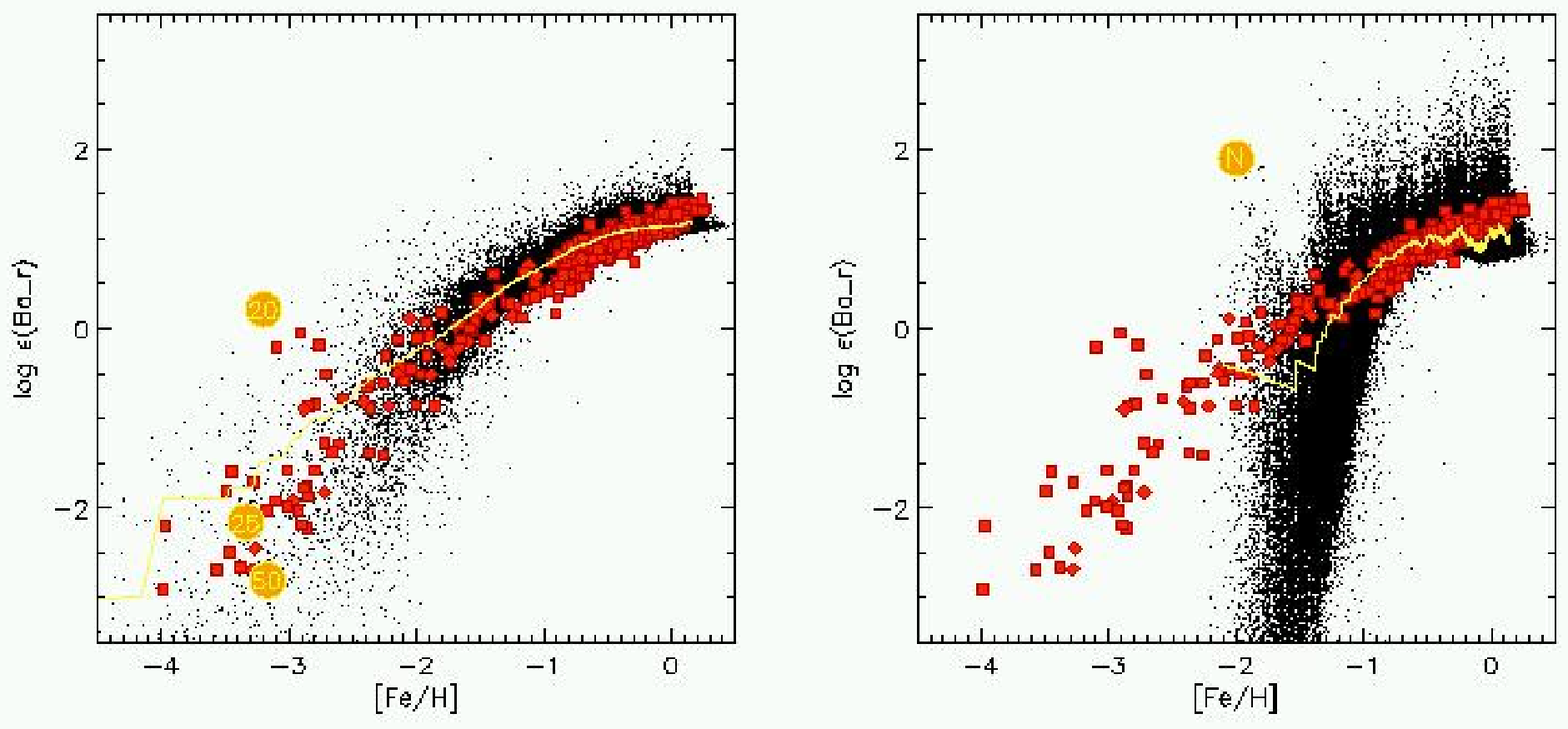}}
 \caption{$\log \epsilon \left( \mathrm{Ba}^r \right)$ vs. [Fe/H] of model
          SN2050 (left panel) and NSM occurring at a rate of $2 \cdot10^{-5}$
          yr$^{-1}$ (right panel).  Symbols are as in Fig.~\ref{sn810.fig}.
          Additionally, numbered circles in the left panel give $\log \epsilon
          \left( \mathrm{Ba}^r \right)$ of SNe~II with the indicated
          progenitor mass, when mixed with $5 \cdot 10^4 \, \mathrm{M}_{\sun}$
          of primordial ISM. Similarly, the labelled circle in the right panel
          indicates $\log \epsilon \left( \mathrm{Ba}^r \right)$ of NSM when
          mixed with $5 \cdot 10^4 \, \mathrm{M}_{\sun}$ of enriched ISM
          (at [Fe/H] $=-2$).}
 \label{loge.fig}
\end{figure*}

The following qualitative differences are immediately visible, if the upper
right panel of Fig.~\ref{nsm001.fig} is compared to
Figs.~\ref{sn810.fig}~--~\ref{sn2050.fig}:
\begin{enumerate}
\item r-process nuclei appear at the earliest around [Fe/H] $\approx -2.5$,
whereas r-process elements such as Eu are observed at [Fe/H] $= -3$ and
probably even down to [Fe/H] $=-4$ in the case of Ba$^r$. The reason for this
late injection of r-process matter in the model is the low NSM rate (compared
to the SN~II rate). The decisive parameter to obtain a given NSM rate in the
model is the probability $P_{\mathrm{NSM}}$ which, in this case, is set to $4
\cdot 10^{-3}$ (see Table~\ref{nsmparam}). Thus, on average $\approx 250$
SNe~II events occur before the first NSM event and r-process nuclei are
injected into an ISM already pre-enriched by SNe~II. The lower the NSM rate,
i.e. the lower $P_{\mathrm{NSM}}$, the later (in time and metallicity) the
occurrence of r-process elements. This late injection of r-process nuclei by
NSM is the reason why we did not consider coalescence timescales of the order
of 1 Gyr: In our model, 1 Gyr after the first SF event the metallicity of the
ISM reached [Fe/H] $\approx -0.9$. The first appearance of r-process nuclei at
this metallicity is clearly not consistent with observations.  Thus, the
advantage that the NSM scenario can produce a large scatter in [r/Fe] close to
3 dex, as observed in ultra metal-poor stars, turns into a disadvantage
because it appears at too high metallicities where observations do not show
this effect anymore.
\item There is a prominent tail of model stars with very low [r/Fe] ratios at
[Fe/H] $= -2$. Such model stars are also present in
Figs.~\ref{sn810.fig}~--~\ref{sn2050.fig}. In this case, however, a
significant fraction of them shows this property. Furthermore, the tail
develops not until [Fe/H] $\ge -2$, whereas in the SN~II case there are
virtually no model stars with very low [r/Fe] ratios above this
metallicity. Since the tail develops at higher metallicities, it can not be
dismissed as unobservable. Even if r-process element abundances with [r/Fe]
$\le -1$ could not be detected, the upper part of the tail should be visible
in the observational data.
\item Even at late times in the enrichment of the ISM ([Fe/H] $\ge -1$), the
scatter in possible [r/Fe] ratios is of the order $1.5-2.0$ dex, whereas
observations of [Eu/Fe] and [Ba$^r$/Fe] abundances show a scatter of
approximately $0.2-0.3$ dex. The large scatter at high [Fe/H] occurs only for
NSM but not for SNe~II. Since the total amount of r-process matter ejected in
a single NSM event (here $10^{-3} \, \mathrm{M}_{\sun}$) is very large (cf. to
$\approx 10^{-6}-10^{-5} \, \mathrm{M}_{\sun}$ for SNe~II) and the frequency
of NSM is very low ($\approx 100 - 1000$ times lower than that of SNe~II), NSM
events may still cause significant local chemical inhomogeneities, in spite of
the advanced enrichment of the ISM by many SNe~II.
\end{enumerate}

These aspects strongly argue against NSM as the \emph{dominating} r-process
source, especially since the parameters used for the model in discussion are
at the upper limit set by theoretical and observational constraints. Lower NSM
rates and, consequently, larger ejecta masses, only aggravate the problems
mentioned above. The dramatic changes in the distribution of r-process
abundances occurring with decreasing NSM rate are clearly visible in the
sequence of plots in Figs.~\ref{nsm001.fig} and \ref{nsm100.fig}. The NSM
rates adopted in the models decrease from left to right and top to bottom by
one order of magnitude for each panel, i.e. from $2 \cdot 10^{-3} \,
\mathrm{yr}^{-1}$ for the upper left panel down to $2 \cdot 10^{-6} \,
\mathrm{yr}^{-1}$ for the lower right panel.  Figs.~\ref{nsm001.fig} and
\ref{nsm100.fig} assume coalescence timescales of 1 and 100 Myr, respectively.

In the upper left panels of both figures (NSM rate $2 \cdot 10^{-3} \,
\mathrm{yr}^{-1}$), a few model stars with r-process abundances first appear
around [Fe/H] $\approx -3.5$, in contrast to [Fe/H] $\approx -1$ in the panels
at the lower right (NSM rate $2 \cdot 10^{-6} \, \mathrm{yr}^{-1}$).
Simultaneously, the scatter in [Ba$^r$/Fe] and [Eu/Fe] at solar metallicity,
which is of the same order as the one observed in the upper left panels,
increases to almost 3 dex in the lower right panels, clearly not consistent
with observations. Furthermore, the tail of model stars with low [r/Fe]
abundances gets more pronounced and concentrated and is shifted to higher
metallicities.

Contrary to expectations, the impact of the coalescence timescale $t_c$ on the
distribution of r-process nuclei is rather small. This is surprising, since a
large value of $t_c$ results in a delayed injection of r-process matter into
the ISM. However, this can be understood by noting that in the NSM models, the
enrichment of the ISM was adjusted to proceed very slowly at the beginning of
Galaxy formation: It takes $\approx 200$ Myr to enrich the ISM from [Fe/H] $=
-4$ to [Fe/H] $= -3$. Thus, a coalescence timescale of 100 Myr is of the
same order as the halo enrichment timescale, resulting in only a slight shift
of the first r-process matter injection to higher metallicities. The slow
enrichment was assumed in order to enable r-process nuclei injection from NSM
even at low metallicities. However, as Figs.~\ref{nsm001.fig} and
\ref{nsm100.fig} show, this is not the case even with the help of a slow
enrichment history. Thus, a fast halo enrichment would shift the moment of
first r-process matter injection to even higher metallicities. Note that, on
the basis of these considerations, coalescence timescales of the order 1 Gyr
clearly are not consistent with the occurrence of r-process nuclei early on
during Galaxy formation.

Fig.~\ref{loge.fig} demonstrates another qualitative difference between the
SN~II and NSM scenarios. In this figure, $\log \epsilon \left( \mathrm{Ba}^r
\right)$\footnote{$\log \epsilon \left( A \right) = \log \left( N_A/N_H
\right) + 12$} vs. [Fe/H] of model SN2050 (left panel) and NSM occurring at a
rate of $2 \cdot10^{-5}$ yr$^{-1}$ (right panel) are shown. In the SN~II
scenario, the scatter in $\log \epsilon \left( \mathrm{Ba}^r \right)$ at low
metallicities ([Fe/H] $< -2$) is primarily determined by the progenitor mass
dependent r-process yields of SNe~II (indicated by numbered circles). Model
stars at [Fe/H] $< -2$ with Ba$^r$ abundances above $\log \epsilon \left(
\mathrm{Ba}^r \right) = -3$ are formed out of material enriched by a single
SN~II whereas some model stars with $\log \epsilon \left( \mathrm{Ba}^r
\right) < -3$ are formed out of material which underwent subsequent mixing
with primordial ISM. Stars with such low r-process abundances disappear at
[Fe/H] $> -2$ due to the increased enrichment of the ISM with r-process
material.

In the NSM case, the scatter in $\log \epsilon \left( \mathrm{Ba}^r \right)$
is solely determined by subsequent mixing of r-process enriched material with
r-process poor ISM. Under the assumption that a NSM event is able to enrich $5
\cdot 10^4 \, \mathrm{M}_{\sun}$ of ISM with $10^{-2} \, \mathrm{M}_{\sun}$ of
r-process matter, model stars with typically $\log \epsilon \left(
\mathrm{Ba}^r \right) = 2$ or more can be formed (indicated by the labelled
circle in the right-hand panel of Fig.~\ref{loge.fig}). Thus, the distinct
peaks in the distribution of Ba$^r$ abundances of model stars are telltale
signs of recent NSM events.  However, since NSM events are rare, there is time
enough to thoroughly mix the r-process enriched material in the vicinity of a
recent NSM with r-process poor ISM. In consequence of this mixing and the low
NSM rate, the bulk of model stars with $\log \epsilon \left( \mathrm{Ba}^r
\right) < 2$ are formed.

We conclude that NSM have to respect stringent requirements if they are to
operate as dominating r-process contributors, which most likely are not
fulfilled in reality: High NSM rates of the order $\ge 10^{-3} \,
\mathrm{yr}^{-1}$ are required to account for the early appearance of
r-process nuclei in the ISM. Consequently, the amount of r-process matter
ejected in a NSM event has to be of the order $\le 10^{-4} \,
\mathrm{M}_{\sun}$, lest the total inventory of r-process nuclei in the Galaxy
is exceeded. In addition, the coalescence timescale of a large fraction of
neutron star binaries has to be of the order 1-10 Myr or at the most 100
Myr. If these constraints are indeed fulfilled in reality, then the properties
of NSM (i.e. rates and ejected r-process matter) are comparable to the ones
of SNe~II that (may) give rise to r-process nucleosynthesis from the point of
view of chemical evolution. Consequently, the enrichment of the ISM with
r-process matter from NSM would be similar to the one where core-collapse SNe
are the source of r-process elements (compare
Figs.~\ref{sn810.fig}~--~\ref{sn2025.fig} with the upper left panels of
Figs.~\ref{nsm001.fig} and \ref{nsm100.fig}).

However, it seems unlikely that NSM fulfil these conditions, which suggests
that NSM are ruled out as the major r-process source. On the other hand, NSM
occurring at low rates and with low ejecta masses of the order $10^{-4} -
10^{-5} \, \mathrm{M}_{\sun}$ may still contribute to the enrichment of the
ISM with r-process nuclei if their r-process signature is similar to the one
generated by SNe~II. In this case, the impact of NSM on r-process nuclei
enrichment would be negligible compared to the fast injection of r-process
matter by SNe~II.


\section{Conclusions}
\label{con}

In this work, we study the enrichment of the interstellar medium with
r-process elements in the framework of inhomogeneous chemical evolution. We
present a detailed comparison of the impact of lower-mass SNe~II ($8-10 \,
\mathrm{M}_{\sun}$, model SN810), higher-mass SNe~II ($\ge 20 \,
\mathrm{M}_{\sun}$, models SN2025 and SN2050) and NSM as major r-process sites
on the enrichment history of the early Galaxy.

Among the SNe~II scenarios, model SN2050 gives the best fit to observations,
since it reproduces the trend of r-process abundances at ultra low
metallicities and at the same time fits observations with the highest [r/Fe]
abundance ratios at [Fe/H] $\approx -3$. Neither model SN2025 nor SN810 are
able to reproduce observations of the most r-process enriched stars. In
addition, model SN810 gives the worst fit to the trend of r-process abundances
at ultra low metallicities.  However, we conclude that, due to the large
uncertainties inherent in the progenitor mass dependence of iron yields of
SNe~II, it is not possible to clearly rule out either lower-mass SNe~II or
higher-mass SNe~II as dominant r-process sites from the point of view of
inhomogeneous chemical evolution.  Additional uncertainties are introduced by
the fact that reliable r-process yields from SNe~II are unavailable as
yet. Here, they were deduced in such a way that the average [r/Fe] abundances
in metal-poor halo stars are reproduced. Clearly, iron and r-process yields
and their dependence on progenitor mass from self-consistent core-collapse SN
calculations are required before any decisive conclusion can be reached.

On the other hand, NSM seem to be ruled out as major r-process sources for the
following reasons:
\begin{enumerate}
\item Estimates of the galactic NSM rate are two to three orders of magnitude
lower than estimates of galactic SNe~II rate. Thus, the injection of r-process
nuclei into the interstellar medium by NSM would occur very late during Galaxy
formation ([Fe/H] $\approx -2.5$), whereas r-process elements are already
observed at [Fe/H] $= -3$ and probably even at [Fe/H] $=-4$.
\item The late injection of r-process elements furthermore leads to prominent
tails in the distribution of r-process abundances down to very low [r/Fe]
ratios at [Fe/H] $\ge -2$, which are seemingly not consistent with
observations. It is possible that r-process abundances [r/Fe] $\le -1$ are not
detectable to date in metal-poor halo stars. However, the large fraction of
stars predicted by our model with $-1 \le$ [r/Fe] $\le 0$ at [Fe/H] $\ge -2$
should be present in the observational data.
\item Since NSM occur at a lower rate than SNe~II, their r-process yield has
to be about two orders of magnitudes higher than the r-process yield of
typical SNe~II. Due to this large r-process yield, considerable chemical
inhomogeneities in the ISM are expected to be present even at solar
metallicity. The scatter in [r/Fe] is predicted to be of the order $2.0 - 2.5$
dex, whereas a scatter of only $0.2 - 0.3$ dex is observed.
\end{enumerate}

NSM as major r-process sources are only consistent with observations under the
following conditions: First, they occur at rates about one to two orders of
magnitude higher than given by present estimates. Second, a \emph{dominant}
part of the NSM population has coalescence timescales shorter than
approximately 10 Myr. Third, the total amount of ejected r-process matter is
of the same order as the present theoretical estimate from relativistic merger
calculations.  Under these conditions, the enrichment of the ISM with
r-process elements from NSM is qualitatively very similar to the enrichment of
r-process elements dominated by SNe~II. However, it seems highly unlikely that
NSM fulfil these conditions, which suggests that NSM are ruled out as the
major r-process source. Nevertheless, it is still feasible from the point of
view of chemical evolution that NSM contribute a minor fraction of r-process
elements to the enrichment of the ISM. In this case they have to occur at a
low rate and the ejected r-process matter must not exceed the lower limit
indicated by recent relativistic merger calculations.

Thus, we conclude, that the \emph{exact} astrophysical nature of r-process
sites still remains a mystery, since it is not possible to clearly distinguish
between neutron capture element abundance patterns resulting from lower-mass
SNe~II ($8-10 \, \mathrm{M}_{\sun}$) and the ones from higher-mass SNe~II
($\ge 20 \, \mathrm{M}_{\sun}$) in the framework of inhomogeneous chemical
evolution. However, the present investigation suggests that core-collapse SNe
are much more likely to be the \emph{dominant} r-process sites than coalescing
neutron star binaries, which at least reduces the list of possible major
contributors of r-process nuclei to the enrichment of the interstellar medium.
Yet, it remains to be seen how SNe~II can actually produce the required
r-process yields.

\acknowledgements{This work was supported in part by the Swiss National
Science Foundation and by US DOE grants DE-FG02-87ER40328 and
DE-FG02-00ER41149 at Minnesota.}



\begin{thebibliography}{}

\bibitem[1989]{ag89} Anders, E., \& Grevesse, N. 1989, Geochimica et Cosmochimica
Acta, 53, 197

\bibitem[2000]{ao00}
Aoki, W., Norris, J.~E., Ryan, S.~G., et al. 2000, ApJ, 536, L97

\bibitem[2000]{ar00}
Argast, D., Samland, M., Gerhard, O.~E., \& Thielemann, F.-K. 2000, A\&A, 356, 873
(Paper~I)

\bibitem[2002]{ar02}
Argast, D., Samland, M., Thielemann, F.-K., \& Gerhard, O.~E. 2002, A\&A, 388, 842

\bibitem[2002]{bl02}
Belczynski, K., Kalogera, V., \& Bulik, T. 2002, ApJ, 572, 407

\bibitem[1994]{be94}
Beveridge, R.~C., \& Sneden, C. 1994, AJ, 108, 285

\bibitem[2000]{brach00}
Brachwitz, F., Dean, D.~J., Hix, W.~R., et al. 2000, ApJ, 536, 934

\bibitem[1989a]{bru89a}
Bruenn, S.~W. 1989a, ApJ, 340, 955

\bibitem[1989b]{bru89b}
Bruenn, S.~W. 1989b, ApJ, 341, 385

\bibitem[1957]{b2fh}
Burbidge, G.~R., Burbidge, W.~A., Fowler, W.~A., \& Hoyle, F. 1957, Rev. Mod. Phys., 29, 547
(B$^2$FH)

\bibitem[2000]{br00}
Burris, D.~L., Pilachowski, C.~A., Armandroff, T.~E., et al. 2000, ApJ 544, 302

\bibitem[2001]{cm01}
Cameron, A.~G.~W. ApJ, 2001, 562, 456

\bibitem[1993]{cb93}
Charbonnel, C., Meynet, G., Maeder, A., et al. 1993, A\&AS, 101, 415

\bibitem[1991]{co91}
Cowan, J.~J., Thielemann, F.-K., \& Truran, J.~W. 1991, Phys. Rep., 208, 267

\bibitem[2002]{co02}
Cowan, J.~J., Sneden, C., Burles, S., et al. 2002, ApJ, 572, 861

\bibitem[1993]{ed93}
Edvardsson, B., Andersen, J., Gustafsson, B., et al. 1993, A\&A, 275, 101

\bibitem[2002]{fi02}
Fields, B.~D., Truran, J.~W., \& Cowan, J.~J. 2002, ApJ, 575, 845

\bibitem[1993]{fr93}
Fran\c{c}ois, P., Spite, M., \& Spite F. 1993, A\&A, 274, 821

\bibitem[1999a]{fr99a}
Freiburghaus, C., Rembges, F., Rauscher, T., et al. 1999a, ApJ, 516, 381

\bibitem[1999b]{fr99b}
Freiburghaus, C., Rosswog, S., \& Thielemann, F.-K. 1999b, ApJ 525, L121

\bibitem[1999]{fy99}
Fryer, C.~L., Woosley, S.~E., \& Hartmann, D.~H. 1999, ApJ, 526, 152

\bibitem[1991a]{gr91a}
Gratton, R.~G., \& Sneden C. 1991a, A\&A, 241, 501

\bibitem[1991b]{gr91b}
Gratton, R.~G., \& Sneden C. 1991b, A\&A, 287, 927

\bibitem[2002]{vh02}
Hill, V., Plez, B., Cayrel, R., et al. 2002, A\&A, 387, 560

\bibitem[1984]{hi84}
Hillebrandt, W., Nomoto, K., \& Wolff, R.~G. 1984, A\&A, 133, 175

\bibitem[1999]{is99}
Ishimaru, Y., \& Wanajo, S. 1999, ApJ, 511, L33

\bibitem[1999]{iwetal99}
Iwamoto, K., Brachwitz, F., Nomoto, K., et al. 1999, ApJS, 125, 439

\bibitem[1999]{je99}
Jehin, E., Magain, P., Neuforge, C., et al. 1999, A\&A, 341, 241

\bibitem[2000]{ka00}
Kalogera, V., \& Lorimer, D.~R. 2000, ApJ, 530, 890

\bibitem[2002]{ke02}
Koch, A. \& Edvardsson, B. 2002, A\&A, 381, 500

\bibitem[1991]{la91}
Larson, R.~B., 1991. In: Lambert, D.~A. (ed.) Frontiers of stellar evolution.
ASP Conf. Ser. 20, 539.

\bibitem[2001]{li01} 
Liebend\"orfer, M., Mezzacappa, A., Thielemann, F.-K., et al. 2001, Phys. Rev. D,
6310, 3004

\bibitem[2000]{mg00}
Mashonkina, L. \& Gehren, T. 2000, A\&A, 364, 249

\bibitem[2001]{mg01}
Mashonkina, L. \& Gehren, T. 2001, A\&A, 376, 232

\bibitem[1992]{mz92}
Mazzali, P.~A., Lucy, L.~B. \& Buttler, K. 1992, A\&A, 258, 399

\bibitem[1995a]{mw95a}
McWilliam, A., Preston, G.~W., Sneden, C., \& Searle, L. 1995a, AJ, 109, 2757

\bibitem[1995b]{mw95b}
McWilliam, A., Preston, G.~W., Sneden, C., \& Shectman, S. 1995b, AJ, 109, 2736

\bibitem[1998]{mw98}
McWilliam, A. 1998, AJ, 115, 1640

\bibitem[1997]{mb97}
Meyer, B.~S., \& Brown, J.~S. 1997, ApJS, 112, 199

\bibitem[2001]{mi01}
Mishenina, T.~V. \& Kovtyukh, V.~V. 2001, A\&A, 370, 951

\bibitem[2001]{na01}
Nagataki, S. 2001, ApJ, 551, 429

\bibitem[1997]{no97}
Nomoto, K., Hashimoto, M., Tsujimoto, T., et al. 1997, Nucl. Phys., A616, 79c13

\bibitem[2002]{oe02}
Oechslin, R., Rosswog, S., \& Thielemann, F.~K. 2002, Phys. Rev., D65, 103005

\bibitem[1990]{pe90}
Peterson, R.~C., Kurucz, R.~L., \& Carney, B.~W. 1990, ApJ, 350, 173

\bibitem[2001]{pf01}
Pfeiffer, B., Kratz, K.-L., Thielemann, F.-K., \& Walters, W. B. 2001,
Nucl. Phys., A693, 282

\bibitem[1998]{pr98}
Portegies Zwart, S.~F., \& Yungel'son, L.~R. 1998, A\&A, 332, 173

\bibitem[2000]{qi00}
Qian, Y.-Z. 2000, ApJ, 534, L67

\bibitem[2001]{qi01}
Qian, Y.-Z. 2001, ApJ, 552, L117

\bibitem[2001]{qiwa01}
Qian, Y.-Z. \& Wasserburg, G.~J. 2001, ApJ, 552, L55

\bibitem[2002]{qiwa02}
Qian, Y.-Z. \& Wasserburg, G.~J. 2002, ApJ, 567, 515

\bibitem[2002]{qiwa03}
Qian, Y.-Z. \& Wasserburg, G.~J. 2003, ApJ, 588, 1099

\bibitem[1996]{qi96}
Qian, Y.-Z., \& Woosley, S.~E. 1996, ApJ, 471, 331

\bibitem[1999]{ri99}
Raiteri, C.~M., Villata, M., Gallino, R., et al. 1999, ApJ, 518, L91

\bibitem[2000]{ra00}
Rampp, M., \& Janka, H.-T. 2000, ApJ, 539, L33

\bibitem[1999]{ro99}
Rosswog, S., Liebend\"orfer, M., Thielemann, F.-K., et al. 1999, A\&A, 341, 499

\bibitem[2000]{ro00}
Rosswog, S., Davies, M.~B., Thielemann, F.-K., \& Piran, T. 2000, A\&A, 360, 171

\bibitem[2002]{ro02}
Rosswog, S., \& Davies, M.~B. 2002, MNRAS, 334, 481

\bibitem[1991]{ry91}
Ryan, S.~G., Norris, J.~E., \& Bessell, M.~S. 1991, AJ, 102, 303 

\bibitem[1996]{ry96}
Ryan, S.~G., Norris, J.~E., \& Beers, T.~C. 1996, ApJ, 471, 254

\bibitem[1998]{sa98}
Samland, M. 1998, ApJ, 496, 155

\bibitem[1993a]{sr93a}
Schaerer, D., Meynet, G., Maeder, A., \& Schaller, G., 1993a, A\&AS, 98, 523

\bibitem[1993b]{sr93b}
Schaerer, D., Charbonnel, C., Meynet, G., et al. 1993b, A\&AS, 102, 339

\bibitem[1992]{sl92}
Schaller, G., Schaerer, D., Meynet, G., \& Maeder, A., 1992 A\&AS, 96, 269

\bibitem[1959]{sch59}
Schmidt, M. 1959, ApJ, 121, 161

\bibitem[1998]{sh98}
Shigeyama, T., \& Tsujimoto, T. 1998, ApJ, 507, L135

\bibitem[2000a]{sn00}
Sneden, C., Cowan, J.~J., Ivans, I.~I., et al. 2000, ApJ, 533, L139

\bibitem[2002]{st02}
Stephens, A., \& Boesgaard, A.~M. 2002, AJ, 123, 1647

\bibitem[2001]{sm01}
Sumiyoshi, K., Terasawa, M., Mathews, G.~J., et al. 2001, ApJ, 562, 880

\bibitem[1994]{tk94}
Takahashi, K., Witti, J., \& Janka, H.-T. 1994, A\&A, 286, 857

\bibitem[1994]{tm94}
Tammann, G.~A., L\"offler, W., \& Schr\"oder, A. 1994, ApJS, 92, 487

\bibitem[2002]{te02}
Terasawa, M., Sumiyoshi, K., Yamada, S., et al. 2002, ApJ, 578, L137

\bibitem[1996]{th96}
Thielemann, F.-K., Nomoto, K., \& Hashimoto, M. 1996, ApJ, 460, 408

\bibitem[2002]{th02}
Thielemann, F.-K., Hauser, P., Kolbe, E., et al. 2002, SSRv, 100, 277

\bibitem[2001]{tp01}
Thompson, T.~A., Burrows, A., \& Bradley, S.~M. 2001, ApJ, 562, 887

\bibitem[2003]{tp03}
Thompson, T.~A. 2003, ApJ, 585, L33

\bibitem[1999]{tr99}
Travaglio, C., Galli, D., Gallino, R., et al. 1999, ApJ, 521, 691

\bibitem[2001]{tr01}
Travaglio, C., Galli, D. \& Burkert, A. 2001, ApJ, 547, 217

\bibitem[1999]{ts99}
Tsujimoto, T., Shigeyama, T., \& Yoshii, Y. 1999, ApJ, 519, L63

\bibitem[2000]{ts00}
Tsujimoto, T., Shigeyama, T., \& Yoshii, Y. 2000, ApJ, 531, L33

\bibitem[2001]{ts01}
Tsujimoto, T., \& Shigeyama, T. 2001, ApJ, 561, L97

\bibitem[2002]{ut02}
Utrobin, V.~P., \& Chugai, N.~N. 2002, In: Hillebrandt, W., M\"uller, E. (Eds.)
Proceedings of the 11th Workshop on ``Nuclear Astrophysics''. Max-Planck-Institut
f\"ur Astrophysik, Garching, p.\ 136

\bibitem[1996]{vdH96}
van den Heuvel, E., \& Lorimer, D. 1996, MNRAS, 283, L37

\bibitem[1997]{wl97}
Wallerstein, G., Iben, I., Jr., Parker, P., et al. 1997, Rev. Mod. Phys., 69, 995

\bibitem[2001]{wn01}
Wanajo, S., Toshitaka, K., Mathews, G.~J., \& Otsuki, K. 2001, ApJ 554, 578

\bibitem[2002]{wn02}
Wanajo, S., Naoki, I., Ishimaru, Y. et al. 2002, ApJ 577, 853

\bibitem[1996]{wa96}
Wasserburg, G.~J., Busso, M., \& Gallino, R. 1996, ApJ, 466, L109

\bibitem[2000]{wa00}
Wasserburg, G.~J., \& Qian, Y.-Z. 2000, ApJ, 529, L21

\bibitem[2000]{we00}
Westin, J., Sneden, C., Gustafsson, B., \& Cowan, J.~J. 2000, ApJ, 530, 783

\bibitem[1994]{wi94}
Witti, J., Janka, H.-T., \& Takahashi, K. 1994, A\&A, 286, 841

\bibitem[1998]{wh98}
Wheeler, J.~C., Cowan, J.~J., \& Hillebrandt, W. 1998, ApJ, 493, L101

\bibitem[1987]{wi87}
Williams, R.~E. 1987, ApJ, 320, L117

\bibitem[1992]{wo92}
Woosley, S.~E., \& Hoffman, R.~D.  1992, ApJ, 395, 202

\bibitem[1994]{wo94}
Woosley, S.~E., Wilson, J.~R., Mathews, G.~J., et al. 1994, ApJ, 433, 229

\end{thebibliography}
\end{document}